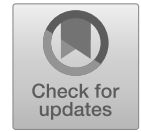

# The Modular X- and Gamma-Ray Sensor (MXGS) of the ASIM Payload on the International Space Station

Nikolai Østgaard[1] · Jan E. Balling[2] · Thomas Bjørnsen[1] · Peter Brauer[2] ·
Carl Budtz-Jørgensen[2] · Waldemar Bujwan[3] · Brant Carlson[1] · Freddy Christiansen[2] ·
Paul Connell[4] · Chris Eyles[4] · Dominik Fehlker[1] · Georgi Genov[1] · Pawel Grudziński[3] ·
Pavlo Kochkin[1] · Anja Kohfeldt[1] · Irfan Kuvvetli[2] · Per Lundahl Thomsen[2] ·
Søren Møller Pedersen[2] · Javier Navarro-Gonzalez[4] · Torsten Neubert[2] ·
Kåre Njøten[1] · Piotr Orleanski[3] · Bilal Hasan Qureshi[1] ·
Linga Reddy Cenkeramaddi[1] · Victor Reglero[4] · Manuel Reina[5] ·
Juan Manuel Rodrigo[4] · Maja Rostad[1] · Maria D. Sabau[5] ·
Steen Savstrup Kristensen[2] · Yngve Skogseide[1] · Arne Solberg[1] · Johan Stadsnes[1] ·
Kjetil Ullaland[1] · Shiming Yang[1]

**Abstract** The Modular X- and Gamma-ray Sensor (MXGS) is an imaging and spectral X- and Gamma-ray instrument mounted on the starboard side of the Columbus module on the International Space Station. Together with the Modular Multi-Spectral Imaging Assembly (MMIA) (Chanrion et al. this issue) MXGS constitutes the instruments of the Atmosphere-Space Interactions Monitor (ASIM) (Neubert et al. this issue). The main objectives of MXGS are to image and measure the spectrum of X- and $\gamma$-rays from lightning discharges, known as Terrestrial Gamma-ray Flashes (TGFs), and for MMIA to image and perform high speed photometry of Transient Luminous Events (TLEs) and lightning discharges. With these two instruments specifically designed to explore the relation between electrical discharges, TLEs and TGFs, ASIM is the first mission of its kind. With an imaging system and a large detector area MXGS will, for the first time, allow estimation of the location of the source region and characterization of the energy spectrum of individual events. The sensors have fast readout electronics to minimize pileup effects, giving high time resolution of photon detection for comparison with measurements on µs-time scales of lightning processes measured by the MMIA and other sensors in space or on the ground. The detectors cover the large energy range of the relevant photon energies. In this paper we describe the scientific objectives, design, performance, imaging capabilities and operational modes of the MXGS instrument.

# 1 Introduction

TGFs were first observed in 1991 by the Burst and Transient Source Experiment (BATSE) onboard the Compton Gamma-Ray Observatory and reported a few years later by Fishman et al. (1994). At the time, the duration of TGFs were thought to be ∼1 ms and photon energies beyond 1 MeV. They were observed when the satellite was passing over regions of active thunderstorms and were at that time thought to originate from altitudes above ∼40 km. With new measurements obtained by the Reuven Ramaty High Energy Solar Spectroscopic Imager (RHESSI) it was shown that the average duration was a few hundreds of microseconds and with energies up to at least 20 MeV, and that the spectral shape pointed to a production altitude below 20 km (Dwyer and Smith 2005). Energies up to 40 MeV was later measured by the Italian satellite, Astrorivelatore Gamma a Immagini LEggaro (AGILE) (Marisaldi et al. 2010), and duration down to just a few tens of microseconds was determined by the Gamma Burst Monitor (GBM) on the Fermi satellite (Connaughton et al. 2013). Two other missions have also reported measurements of TGFs; Beppo-Sax (Ursi et al. 2017) and RELEC (Bogomolov et al. 2017). It should be mentioned that none of these missions were specifically designed to detect the very short flashes of high $\gamma$-ray fluxes, typical for TGFs. Consequently, dead-time and pile-up issues have been a challenge in analyzing the data. For the design of MXGS we have therefore paid special attention to having detector response as well as read-out electronics fast enough to overcome these problems. Having high time resolution (microsecond level) is also important for comparing the $\gamma$-ray measurements with the fast optical pulses from lightning provided by MMIA as well as all measurements from other platforms or from ground of the fast microsecond processes in lightning.

Another lesson learned from earlier measurements of TGFs is the importance of detecting photons in the energy range from the lowest energies around 20 keV up to at least 20 MeV. This is achieved with two co-planar detector layers, a High-Energy Detector (HED) behind a Low-Energy Detector (LED). For the MXGS design these two factors, namely fast detectors/electronics and a large energy range have been important overall requirements. A third driving factor for the design of the MXGS has been to have as large detection area as the platform allows for, both in the low energy range and in the high energy range, in order to obtain the most comprehensive global survey of TGFs to date. A fourth driving factor is to be able to image TGFs with the highest possible resolution over a large field of view (FOV). This means to optimize the pixelated low-energy detector plane and coded mask geometry, as well as shielding this detector plane on all sides except through the coded mask.

The MXGS (Fig. 1) is the first instrument in space specifically designed to measure Terrestrial Gamma-ray Flashes (TGFs).

The structure of the paper is as follows:

- Science Objectives
- MXGS Instrument Overview
- MXGS Sensitivity and Performance
- MXGS On-Board Software and Data Processing
- Summary

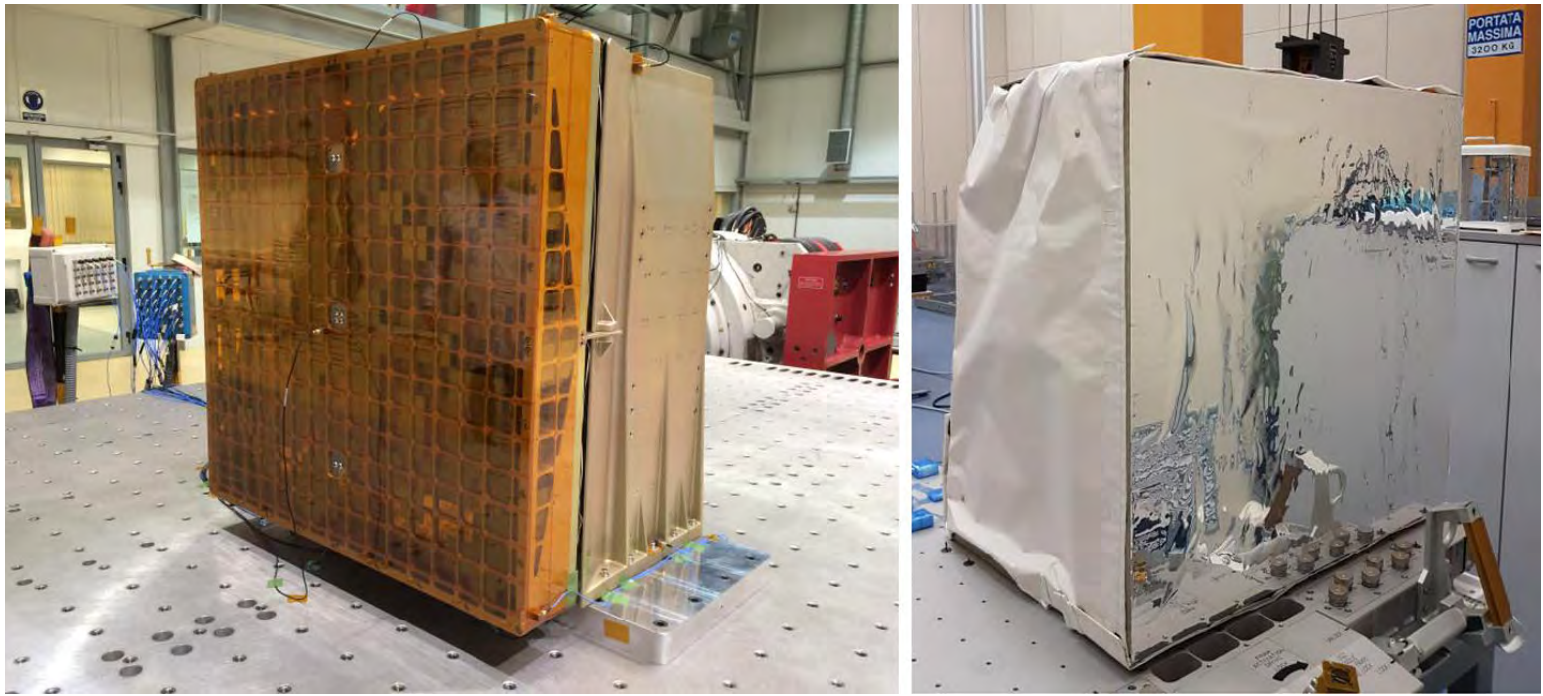

**Fig. 1** The fully assembled MXGS flight model before (left) and after (right) it is covered by thermal blankets and foil and mounted onto the Columbus External Payload Adapter (CEPA) platform. The mass is 147 kg and external dimensions are 774.9 × 746.0 × 523.6 mm

**Table 1** Science objectives of the ASIM mission

---

Research objectives related to the Atmosphere and Climate:
- *Provide the most comprehensive global survey of TLEs and TGFs*
- *Study the physics of TLEs and TGFs*
- *Study how TLEs and TGFs are related to lightning*
- Quantify effects of gravity waves on the mesosphere
- Study high-altitude cloud formation
- Determine the characteristics of thunderstorms that make them effective perturbers of the high-altitude atmosphere

Research objectives related to Space Science:
- *Study effects of thunderstorms on the ionosphere and the radiation belts*
- Determine the distribution of meteors in the Earth's environment and quantify their effect on the atmosphere
- *Understand lightning-induced electron precipitation and relativistic electron precipitation*
- *Study the auroral processes*

Research objectives related to Earth Observation:
- Explore effect of dust storms, mega cities, forest fires and volcanoes on cloud formation and electrification
- Study the relation of eye-wall lightning to the intensification of hurricanes.

---

## 2 Science Objectives

The overall science objectives for the ASIM mission are listed in Table 1. The ones that are directly related to the performance of MXGS are marked with italic font. Some of the objectives require addition supplementary measurements from sensors on other spacecraft and on the ground. Here we will briefly explain the importance of these science objectives and how the MXGS measurements can assist in reaching them.

## 2.1 The Most Comprehensive Global Survey of TGFs

Due to its long trigger window (64 ms) compared to the short duration of TGFs (∼100 μs), BATSE only detected 78 TGFs in 9 years, while RHESSI detected more than 800 TGFs in 6 years. Later, about three times more TGFs have been identified in the RHESSI data (Gjesteland et al. 2012; Østgaard et al. 2015). Both Fermi and AGILE report now about 800 TGF/year (Roberts et al. 2018; Marisaldi et al. 2015). On the other hand, aircraft campaigns flying close to thunderstorm systems have only detected one TGF so far (Smith et al. 2011). An open question is therefore still: How common are TGFs? To establish the occurrence rate of TGFs versus electric discharges in thunderstorms is therefore one of the main scientific goals of MXGS/ASIM. Based on earlier measurements, MXGS sensitivity and ISS altitude we expect to detect ∼1000 TGFs pr year.

## 2.2 The Physics of TGFs and How They Are Related to Lightning

Another outstanding question is to understand how TGFs are produced. We know that X- and $\gamma$-rays in a TGF are produced through the bremsstrahlung process by high energy electrons interacting with the atmosphere, but there is no consensus in the research community on how the electrons are accelerated. Are they produced by electron accelerated in the large-scale ambient fields of thunderstorms or are they produced by electrons accelerated in the local field of lightning leaders? With high relative timing accuracy between optical and $\gamma$-ray measurements, as well as imaging capabilities of both phenomena (lightning and TGFs), ASIM will provide unprecedented measurements to understand the production mechanism of TGFs and the characteristics of the storm regions that produce them. Although the production altitude of TGFs observed from space has been inferred to be between 9 km and 15 km (Cummer et al. 2015, and references therein), it is still an open question what the main production altitude is. This will be addressed in two ways by MXGS. Firstly, the imaging capability should provide a direct measurement and secondly, with its large effective area MXGS will provide energy spectra of single TGFs, thereby also providing information about the production altitude of TGFs. As the imaging capabilities require a certain number of counts (∼32 cnts in the Low Energy Detector, LED) and, due to light contamination, can only be operated during night, we expect to image ∼100 TGFs per year. Light contamination does not affect the High Energy Detector, HED.

## 2.3 Lightning-Induced Electron Precipitation (LEP) and Relativistic Electron Precipitation (REP)

Lightning induced precipitation (LEP) is due to whistler waves produced by lightning, propagating through the ionosphere to interact with high energy particles in the radiation belt. This is a wave-particle interaction that changes the particle distribution such that the particles will increase their parallel velocity component (relative to the magnetic field) at the expense of their normal component and consequently precipitate into the atmosphere and produce X- and $\gamma$-rays through the bremsstrahlung process. Relativistic electron precipitation, which is known to occur during geomagnetic storms, will also produce X- and $\gamma$-rays equatorward of the auroral oval extending many local hours in the magnetic dusk sector. MXGS will have sensitivity adequate to observe LEP (soft) and REP (continuous), see Table 2. The study of LEP and REP are two of the secondary science objectives for MXGS.

**Table 2** X- and $\gamma$-ray source characteristics

| Emission element | Intensity 15–400 keV @ ISS altitude | Intensity 0.2–20 MeV @ ISS altitude | Spectral characteristic | Source altitude [km] | Time scale |
|---|---|---|---|---|---|
| TGF | 0.15 ph/cm$^2$ [a] | 0.35 ph/cm$^2$ [a] | <40 MeV $\sim E^{-1}$ | <20 | $\sim$1–5 ms |
| LEP-soft | $\sim$0.5 ph/cm$^2$/s | – | AO = 4000, efold: 40 keV | >90 | $\sim$1–5 s |
| LEP-hard[a] | $\sim$0.04 ph/cm$^2$/s | – | AO = 100, efold: 85 keV | >90 | $\sim$1–5 s |
| REP-cont | $\sim$55 ph/cm$^2$/s | – | e-folding: 50–200 keV | >100 | Cont. |
| REP-bursts[b] | $\sim$0.01 ph/cm$^2$/event | – | 1/E, from mono-energetic el-beam | >100 | $\sim$0.1 s |
| Aurora | 10000 ph/cm$^2$/s | – | e-folding 2–20 keV | – | Cont. |

[a]Fluence of TGFs follows a power law distribution with exponent of $-2.3$, so only a fraction ($\sim$10%) will have fluence above average and about 30% above the number of counts (32) needed for imaging

[b]Only events with flux 5–10 times the average flux listed for LEP-hard here will be detected

### 2.4 The X-Ray Aurora

With the inclination of ISS of about 51.6° MXGS will occasionally be able to observe the X-ray aurora, with a very high temporal resolution. As the MXGS imaging capabilities are designed for point sources and sources of limited angular extent we will not be able to image the aurora, but with MXGS high time resolution we will be able to study fast pulsations in the aurora. Table 2 summarizes the characteristics of the phenomena MXGS will measure. The values for TGFs are based on modeling an average RHESSI TGF onto the MXGS detector planes at ISS altitude, while the values for LEP are from Rodger et al. (2004) and Clilverd et al. (2004), the REP values from Rodger et al. (2007) and aurora values from Østgaard et al. (2001).

## 3 MXGS Instrument Overview

### 3.1 Design Philosophy and Overview

A conceptual sketch of MXGS is shown in Fig. 2. It comprises a Low Energy Detector (LED) and a High Energy Detector (HED).

The LED detector plane (green in Fig. 2) consists of pixelated Cadmium-Zink-Telluride (CZT) detector crystals that measure photons with energies from 20 to 400 keV. A hopper-shaped collimator defines the 80° × 80° fully coded field of view for the LED. A passive graded shield surrounding the LED housing (1 mm stainless steel, 0.5 mm Sn, 0.25 mm Ta and 2 mm Pb), except top and bottom, shields the LED against the ambient $\gamma$-rays as well as against the fluorescence photons and radioactive decays in surrounding materials. The shield, also used in the Low Energy Gamma Ray Imager, LEGRI on MINISAT-01 (Dean et al. 1991), is efficient for photon energies below 200 keV.

A coded mask provides the imaging capability of the MXGS LED. The mask is a random pattern of square-formed holes mounted on the entry of the hopper shield and photons below 400 keV can only enter the LED through the holes. The mask is covered towards the inside with a Kapton foil which will stop electrons with energies up to 200 keV but will allow photons down to 15 keV to enter the detector. From the penumbra pattern created by the flux of photons from a distant point source, their direction of arrival can be determined. The area of the holes is 50% of the total entry area of the detector. The coded mask is covered towards the outside with a Silver coated Fluorinated Ethylene Propylene foil, which absorbs visible light and UV radiation. The front end of the MXGS LED is covered with a 0.175 mm thick Arlon PCB for contacting with the high voltage supply. The mask structure is also used for support of a weak radioactive source ($^{109}$Cd), which will be used for the in-flight calibration of the LED.

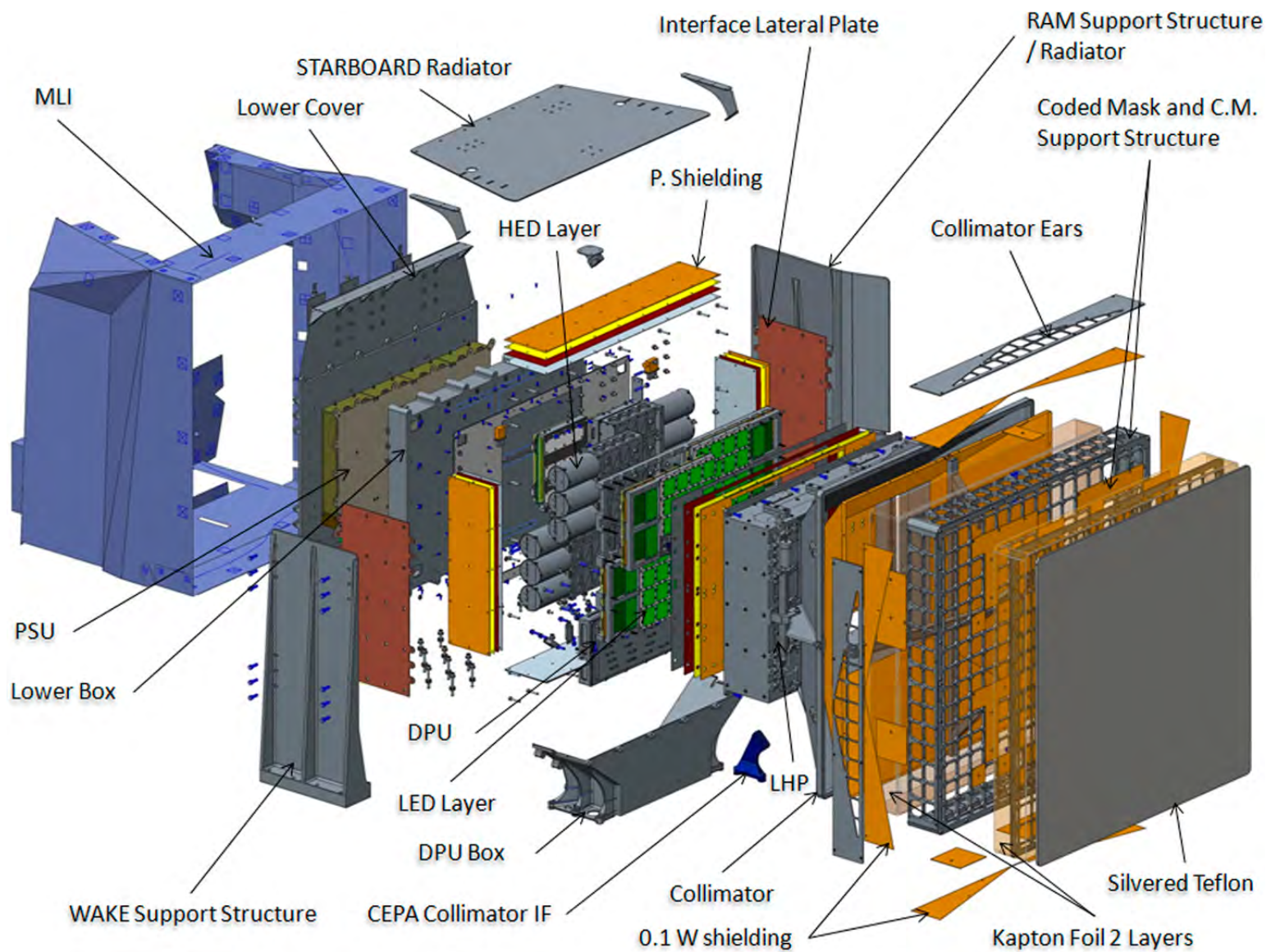

**Fig. 2** Exploded view of MXGS. Main systems

The MXGS HED comprises 12 Bismuth-Germanium-Oxide (BGO) detector bars each coupled to a photomultiplier tube (PMT). HED is sensitive to photons with energies from 200 keV to >20 MeV. The HED is mounted behind the LED and will effectively shield the LED against radiation coming through the rear of the assembly. Three weak $^{22}$Na radioactive sources are mounted in between the CZT detector plane and the BGO array. These sources are used to perform in-flight calibration of the BGO detectors.

The Detector Front End Electronics (DFEE) of the LED is mounted on two sides of its detector plane and for the HED behind the detectors in the housing that contains the HED. Behind the HED housing is the box with Power Supply Unit (PSU). It includes a High Voltage Power Supply (HVPS) and Low Voltage Power Supply (LVPS). The MXGS Data Processing Unit (DPU) is mounted in its own compartment below the electronics box.

Figure 3 shows the functional block diagram and the internal interfaces of MXGS. X and soft $\gamma$-rays, which enter through the opening of the collimator and the coded mask, interact in the CZT crystals of the LED. The energy of the incoming photon is thereby converted to electrical charges. These are drifted to the detector readout electrodes by an electric field, produced in the detector by the LED HVPS. The detector signals are amplified by the pream-plifiers, ASICs, and subsequently digitized and time stamped by the Field-Programmable Gate Array (FPGA) of the LED Detector Assembly Unit (DAU). A large fraction of the $\gamma$-rays with energies above 300 keV will pass through the LED, but will be absorbed in the BGO crystals of the HED. The light produced in this process is read out by the HED PMTs, digitized and processed by the FPGA of the HED DAU. Both the data from LED and HED data are passed on to the DPU, which inspects the data for occurrence of short bursts. A trigger signal will be generated in case of burst detection. The DPU will also format the detector data for telemetry. The full set of parameters for an event consists of arrival time, position in the detector (LED: pixel-ID, HED: BGO-bar) and pulse height (proportional to photon energy). When a burst is detected, all events in a 2-second time interval around the trigger time will be transmitted for further on-ground analysis.

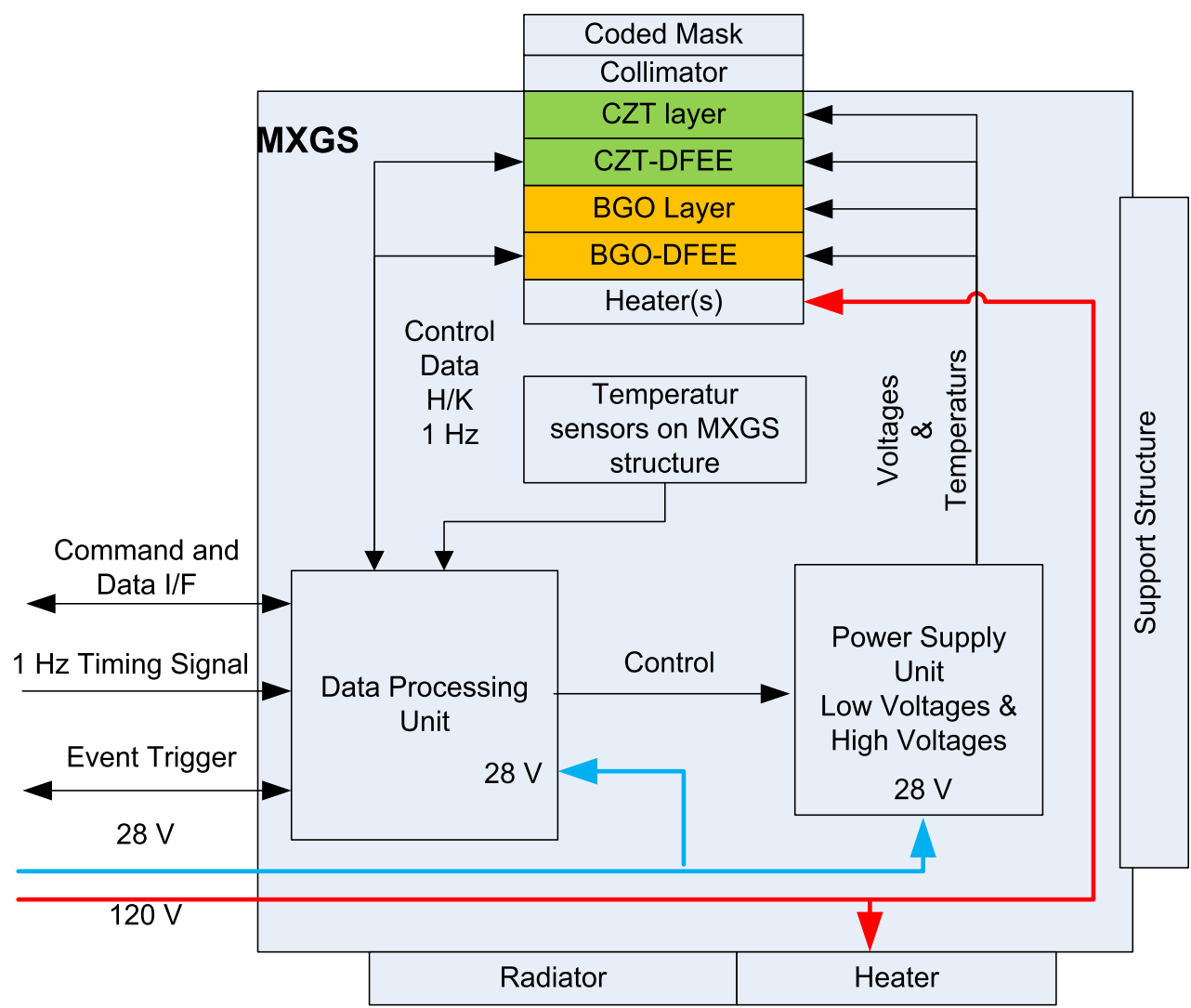

**Fig. 3** MXGS functional block diagram

The LED is a position sensitive pixel detector, which together with the coded mask in front of the LED allows for determining the direction to the source for the detected radiation. The imaging feature of the MXGS is a unique capability, which will be exploited for the first time in connection with TGF research by MXGS. The MXGS is bi-directionally linked to the MMIA instruments. A TLE or a lightning event detected by the MMIA will also cause a 2-second data dump of the MXGS events. MXGS, in order to monitor its performance, will continuously collect spectra of the ambient environment and of the calibration source (see "MXGS On-Board Software and Data Processing". These spectra are transmitted to ground at least once per orbit.

### 3.2 The LED and HED Detectors

The LED and HED are both assembled in four Detector Assembly Units (DAUs).

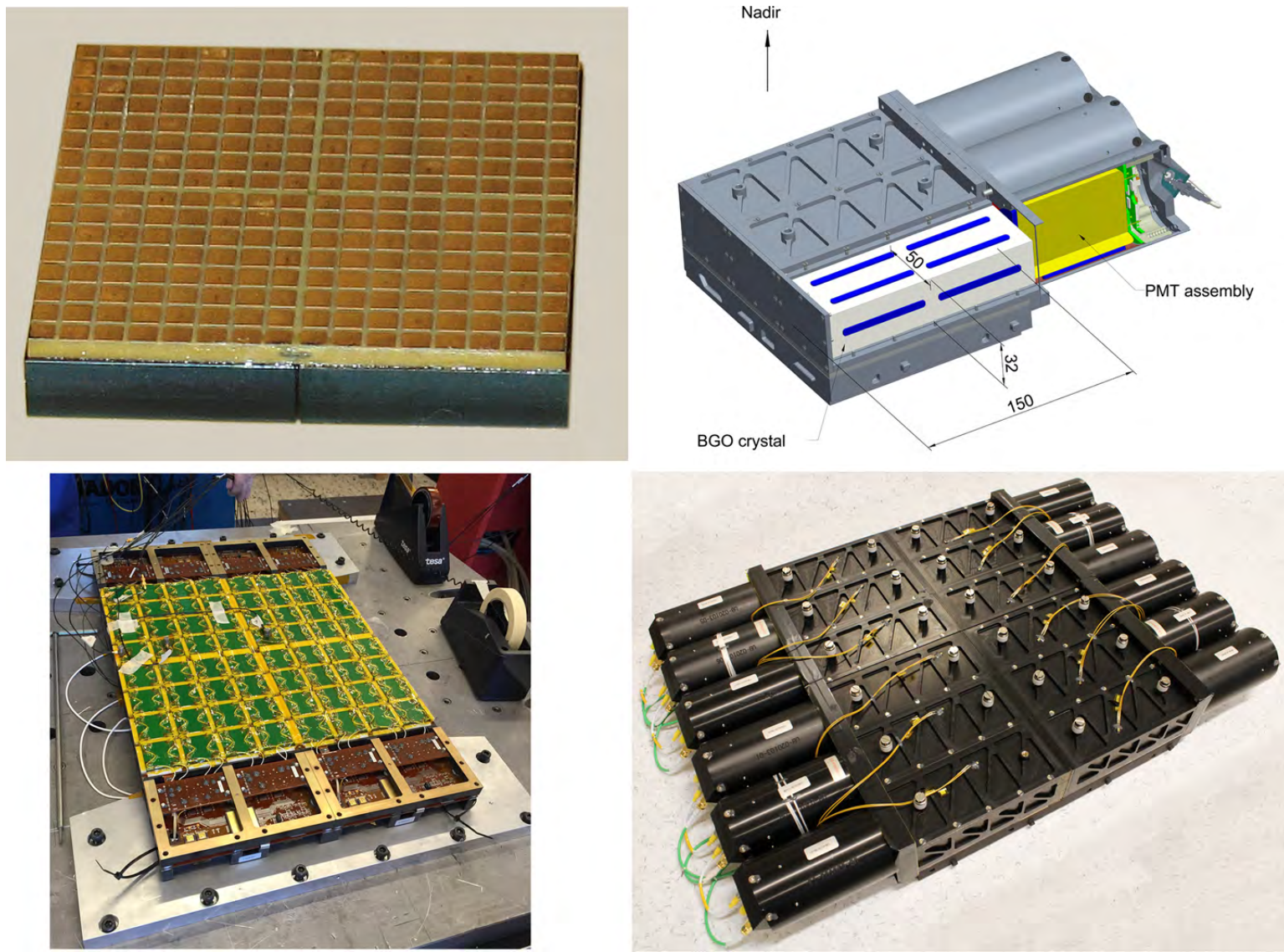

**Fig. 4** Upper left: one detector element of the CZT, $4 \times 4 \times 0.5$ cm$^3$ and $16 \times 16$ pixels. Lower left: LED—all 4 DAUs. Upper right: one HED DAU with 3 PMT/BGO bars. Lower right: HED—all 4 DAUs

The LED detector plane consists of pixelated Cadmium-Zink-Telluride (CZT) detector crystals that measure photons with energies from 20 to 400 keV. CZT is a semiconductor with a high quantum efficiency for hard X- and $\gamma$-rays due to the high atomic numbers of the material. It can be operated close to room temperature (typically $+10$ °C) and does not require cryogenic cooling as required by conventional (thick) Si and Ge detectors. CZT detector systems require modest high voltage, and coupled to ASICs, these systems are compact and ideally suited for space experiments with a low read-out noise and high energy-resolution. Detectors with pixelated readout are well suited for imaging purposes. Each LED DAU consists of 4 chains of 4 Detector Modules (DMs), which are read out separately. Each CZT DM is $4 \times 4 \times 0.5$ cm$^3$, divided into $16 \times 16$ pixels. This gives 4096 pixels per DAU and 16384 for the entire LED. The CZT crystals were provided by REDLEN, Canada. Each CZT DM is integrated with two ASIC-XA-1.82 units, provided by IDEAS, Norway. One DM and the entire LED are shown in Fig. 4, left panels. The total geometrical area of LED is 1024 cm$^2$.

The HED detector plane consists of Bismuth-Germanium-Oxide (BGO) scintillators and is sensitive to photons with energies from 200 keV to $>$20 MeV. BGO has been used in other space missions, as Fermi, for detecting high energy photons. One HED DAU consists of 3 separate BGO bars each one connected to a photomultiplier tube (PMT), which gives 12 BGO/PMT detectors in the HED. The BGO bar is $15 \times 5 \times 3.2$ cm$^3$ and each BGO/PMT is read out separately. The BGO bars were provided by Saint-Gobain and the PMTs (R6231-01 MOD) by Hamamatsu. One HED DAU and the entire HED are shown in Fig. 4 right panels. The total geometrical area of HED is 900 cm$^2$.

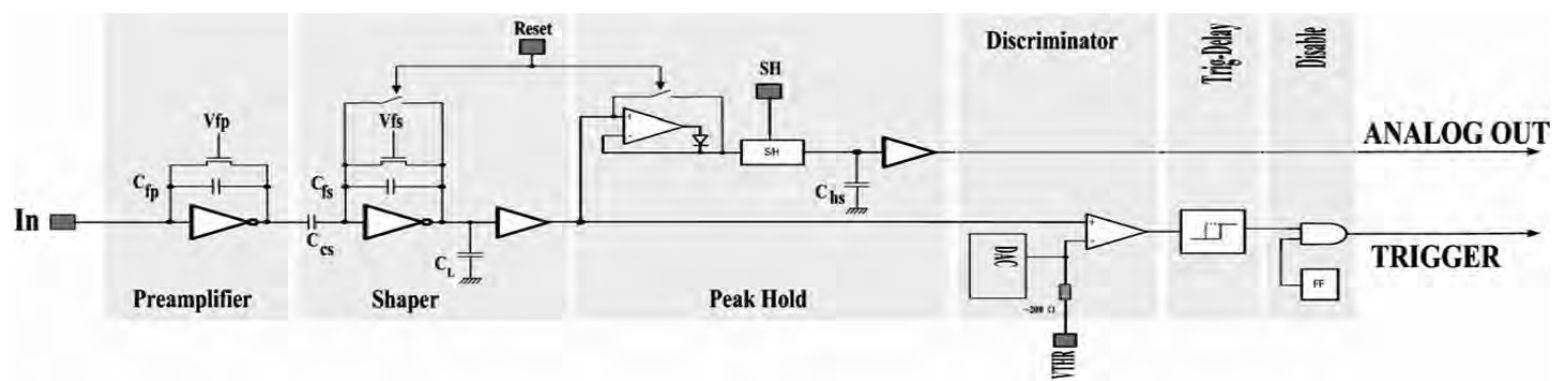

**Fig. 5** Architecture of XA-ASIC channel front-end (from [XA-ASIC manual])

### 3.3 The PSU and the LED and HED Read-Out Electronics

The LED and HED units are self-contained read-out units, each controlled by an anti-fuse FPGA from Microsemi (RTAX 2000). Each unit has a dedicated data and control interface with a bit-rate of 18 Mbit/s connected to the common DPU.

Power is provided by the common Power Supply Unit (PSU), which consists of four identical blocks, each dedicated for powering one LED DAU and one HED DAU. Each PSU block supplies the two DAUs with appropriate LV and HV rails and also provides the PSU controller with analog telemetry channels and command interface. This allows the DPU to turn on and off individual supplies and set high voltage output value and to monitor all output voltages and currents together with temperatures at the critical points of the PSU and the DAUs. One LED DAU consumes about 6.3 W, summing up to 25.3 W for the LED. The HED layer consumes about 7.5 W including HV supply. Including also the consumption of the PSU controller 0.6 W, DPU 5 W and power supply loss, MXGS total power consumption is 70 W.

In order to deliver low-noise voltages to all DAUs (HED and LED) requires careful separation of ground connection to avoid any ground loops and resulting noise and interference. The LED detector is particularly sensitive to ASIC power rail noise and HV ripple. Therefore, the analog ground, the digital ground and the high voltage are galvanic separated. In addition, all power supplies inside the PSU are galvanically isolated from the primary power lines, and the analog, digital and high voltage grounds of each DAU are connected in single point inside the DAU.

The signals from the 64 CZT DMs of the LED are read out by 128 XA-ASICs from IDEAS (two XA-ASICs pr DM). Each ASIC is self-triggering and will give as an output: a trigger signal, the pixel address and its signal amplitude when a gamma photon is detected (see Fig. 5). From the ASICs we get analog energy signals and digital addresses conveyed as current signals, which need to be converted to voltages and digitized. It is possible to connect the ASIC output current signals from several ASICs in a chain, and then the current signals will be added together. In our system 8 ASICs are connected, forming one chain. There are 4 chains in one DAU. For the energy signals a 12-bit ADC (10.5 ENOB) is used per chain for digitization. Finally, we interface to the FPGA where the so-called SCience Data Packages (SCDP) are assembled and forwarded to the DPU. A SCDP is a 48-bit sequence containing a 10-bit event energy, 14-bit pixel address, 20-bit time stamp of 1 μs resolution and various flag bits.

The output of the HED PMTs are voltage signals, which are digitized to 12-bit (10.5 ENOB) at a rate of 36 MHz. Continuous sampling at this rate allows for digital noise filters and advance triggering, permitting for example pile-up handling within the FPGA. A voltage pulse is accepted as a valid event if the voltage value is above a configurable threshold and is captured after peak-detection. A new pulse can be detected approximately 300 ns after peak detection, giving a typical "dead time" per PMT/BGO of 550 ns, since the pulse peaking time is about 250 ns. There is a time-over threshold counter per second available as an instrument housekeeping parameter, allowing detection of paralysation, which may happen if there is a high number of counts within the dead time. The 48-bit HED data packages consists of 12-bit event energy, 2-bit detector address, 27-bit time stamp of 27.8 ns resolution and various flag bits. The read-out stream may contain normal events, pile-up events (labelled fast events, which also contain valley data, and will be explained below) and overflow events. Overflow events are sent if the maximum ADC value is exceeded and energy bits are replaced with duration of the overflow. It is also possible to debug the detectors in-flight by configuring sample data mode, which captures 256 raw data-samples pre-triggered from the detection of an event.

### 3.4 Timing Accuracy, Dead-Time and Pile-Up Effects for LED and HED

Although the MXGS is designed with relatively fast detector response and fast read-out electronics, we cannot exclude dead-time and pile-up effects. For an average TGF, we expect to have about 70 counts in the LED detector and about 200 in the HED detector. Depending on the duration of the $\gamma$-ray flash this may or may not result in pile-up and/or dead time effects.

The temporal resolution for LED is 1 µs given by 20-bit timestamps and for the HED it is given by 27-bit time stamps resulting in a temporal resolution of 28.7 ns.

In the LED the dead-time for each XA-ASIC is 1.4 µs, which means that each XA-ASIC can only record one hit in this interval. In our system 8 XA-ASICs (4 DM) are connected in one chain. If there is a hit in a different XA_ASIC in the same chain within this 1.4 µs window it will be recorded as a multi-hit. This can happen because the chain is not allowed to reset before new hits occur. These events are distinguishable, but their energies are added together and their address is not correct and can therefore not be used for imaging. As long as a new hit occurs on the same chain within 1.4 µs of the previous hit this pile-up effect will continue, resulting in a significantly underestimated recorded count rates for high actual count rates. However, the instrument housekeeping data contains a dedicated counter for total number of hits including multi-hits, therefore a state of paralysation is detectable.

For HED, it is the peaking time of the filtered pulse (250 ns) together with a time window (300 ns) after pulse peak-detection that defines the dead-time of 550 ns. This will also be the systematic delay of the correct time. If more than one hit occurs within this dead-time only one count will be recorded, and the time stamp and energy will be a mixture of the two and dependent on the relationship of the two pulses. This is illustrated in Figs. [6](#)A and [6](#)B. If a new hit occurs after this dead-time of 550 ns it will be recorded with correct time but the energy will be the sum of its energy and the tail of the previous pulse, so-called tail-pile-up. This is illustrated in Fig. [6](#)C. Since we record the valley between these two pulses, it is possible to reconstruct the real pulse height of this tail-piled-up pulse.

Trigger signals between MXGS and MMIA have a resolution of 1 µs and the relative timing accuracy between MMIA and MXGS is 10 µs.

### 3.5 The Data Processing Unit (DPU)

The MXGS DPU (Fig. [7](#)) provides overall control, data acquisition, data processing, telemetry formatting and monitoring functions of the MXGS instrument.

The DPU is based on the radiation hard Xilinx Virtex-5 FX130T FPGA with one LEON3 (soft core) processor. The LEON3 core from Gaisler operates at a clock frequency of

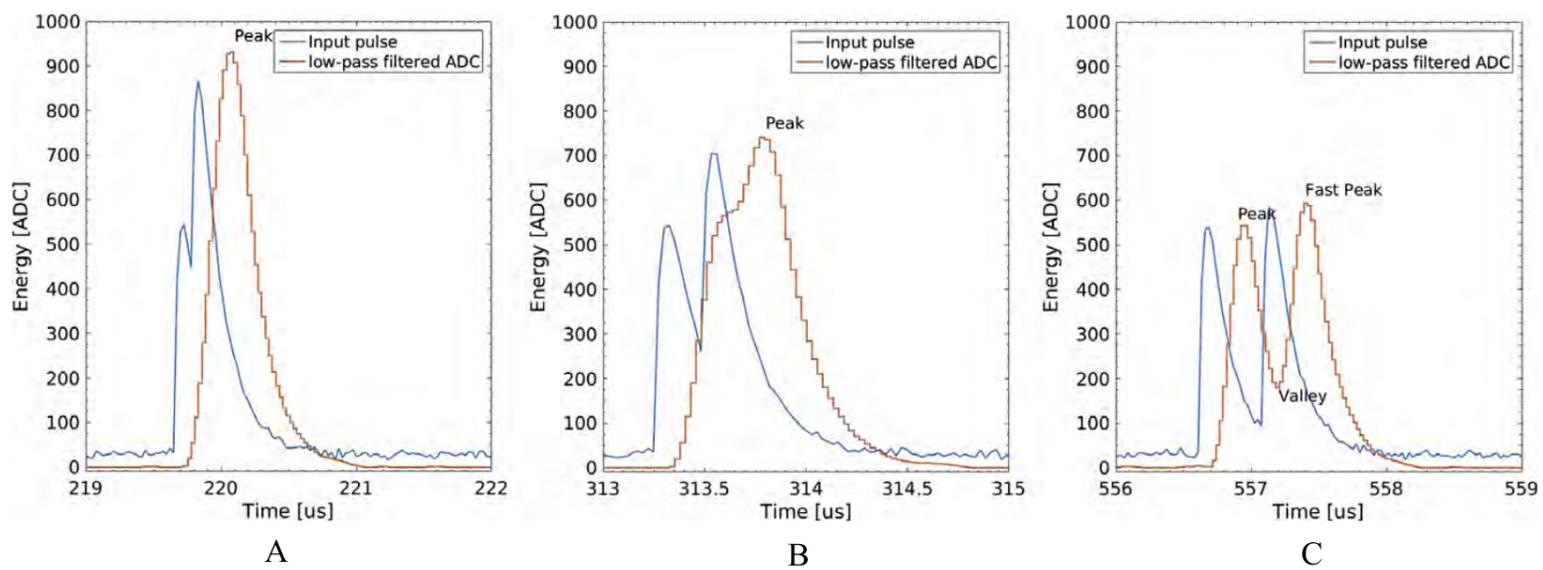

**Fig. 6** Simulated pile-up of BGO/PMT pulse. Blue lines represent simulated input pulse and red lines represent ADC pulses after moving average filter and base-line removal. (**A** and **B**) Two identical pulses, recorded with peak pile-up. Energy and time is a mixture of the two and dependent on pulse height and delay. (**C**) Two identical pulses, the first is correctly recorded, while the second is recorded with tail pile-up

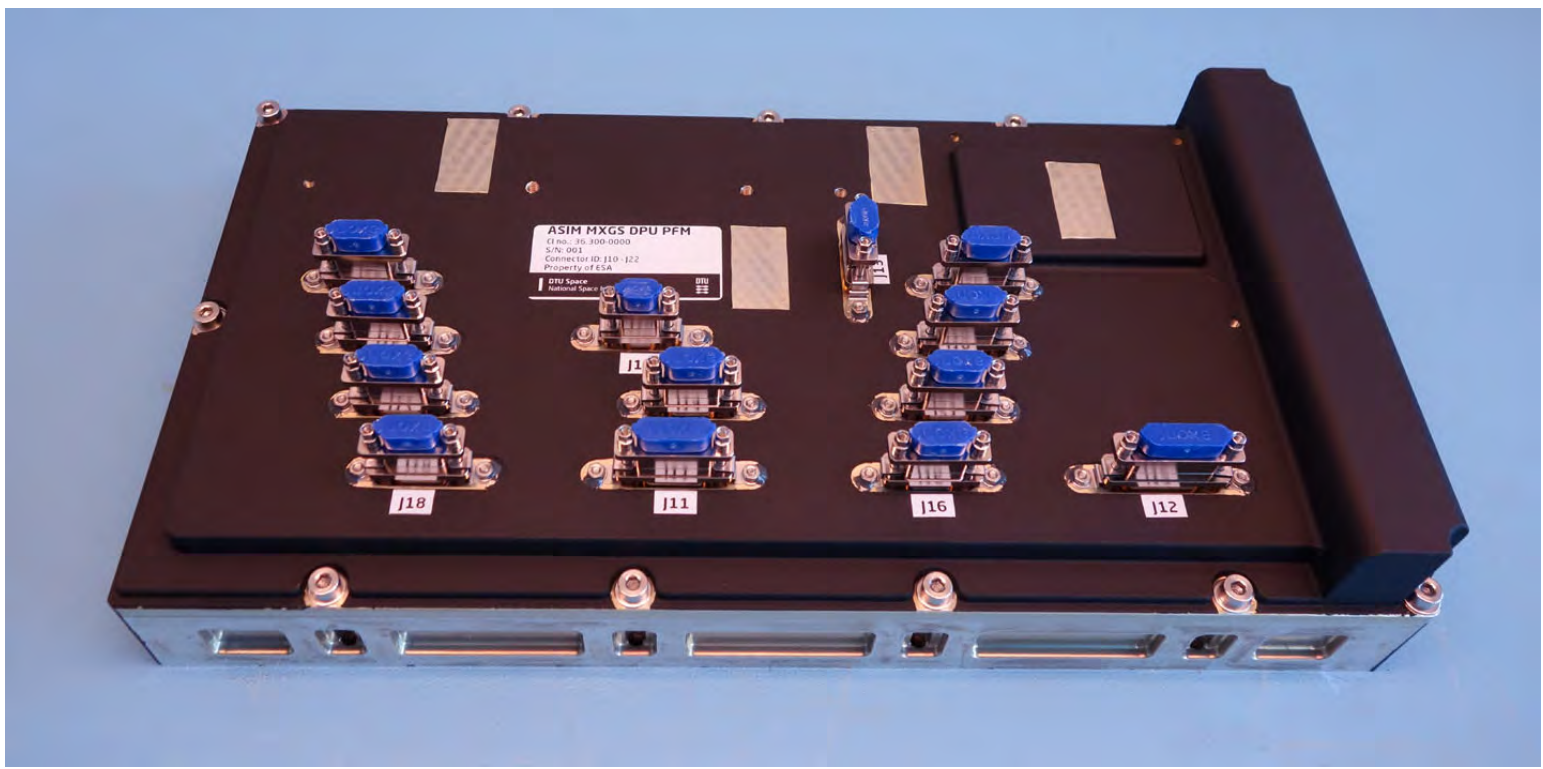

**Fig. 7** MXGS DPU

50 MHz. Integrated into the DPU PCB is an internal power supply which delivers internal voltages of 1.0 V, 2.5 V, 3.0 V and 3.3 V from the 28 V delivered from the Data Handling and Power Unit (DHPU). In Fig. 8 the block diagram of the DPU is shown.

The DPU provides control of LED, HED and PSU. It collects and processes event data from the LED and HED and processes the event data, specifically searching for TGFs. It performs additional event processing related to secondary science objectivities and engineering monitoring (e.g. background event monitoring, detector pulse-height spectra, etc.). With the DPU transmitting and receiving event flags (triggers) between MMIA and MXGS, a TLE or lightning detected by the MMIA will also cause a 2-second data dump of the MXGS events and vice versa.

From the DHPU a 1 Hz Time Correlation Pulse (TCP) is received and encoded onto a DPU generated 1 MHz clock signal that is passed to both LED (CZT) and HED (BGO). This signal is used by LED and HED to time-tag the data with a resolution of 1 μs and combined

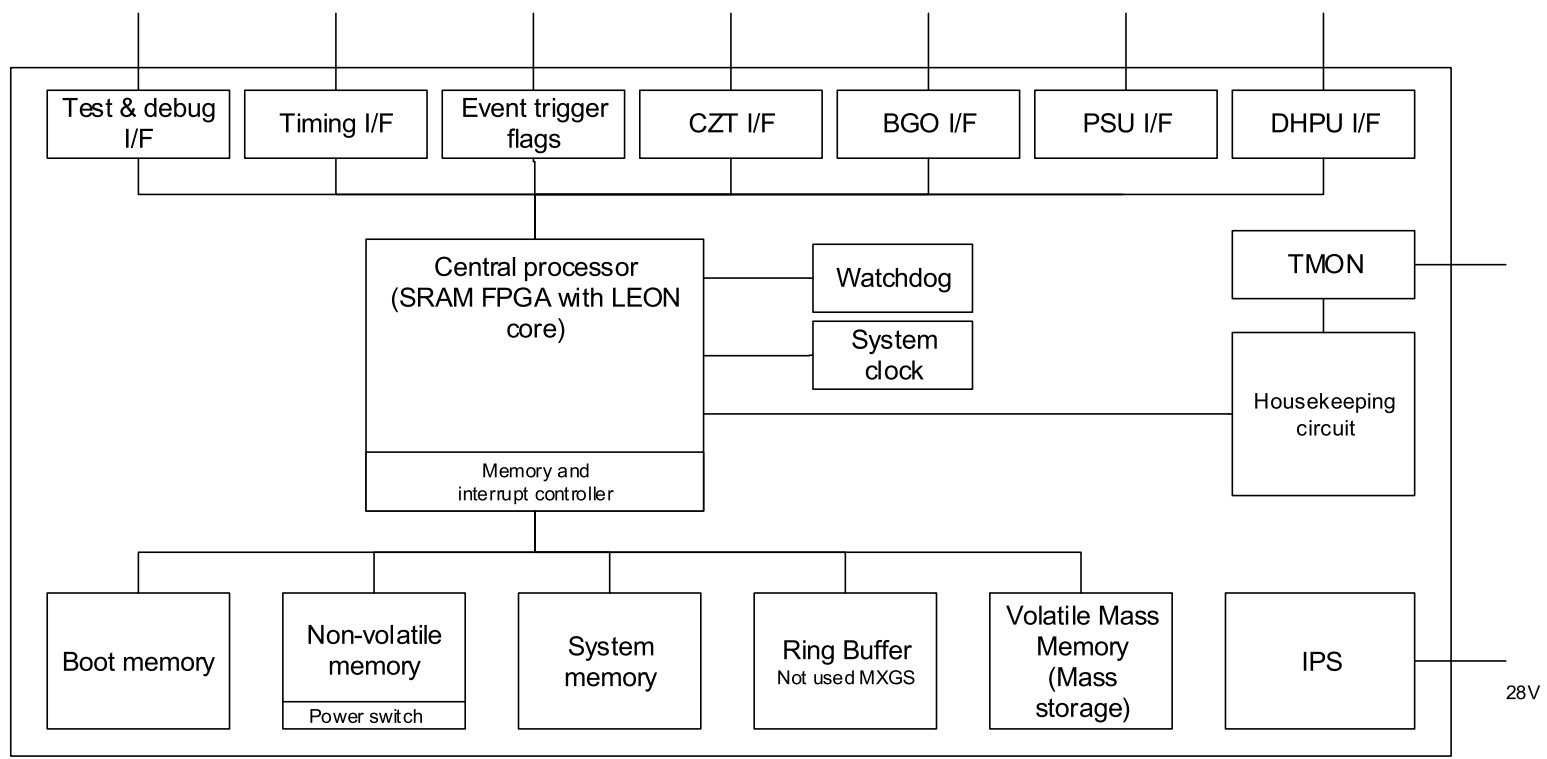

**Fig. 8** MXGS DPU block diagram

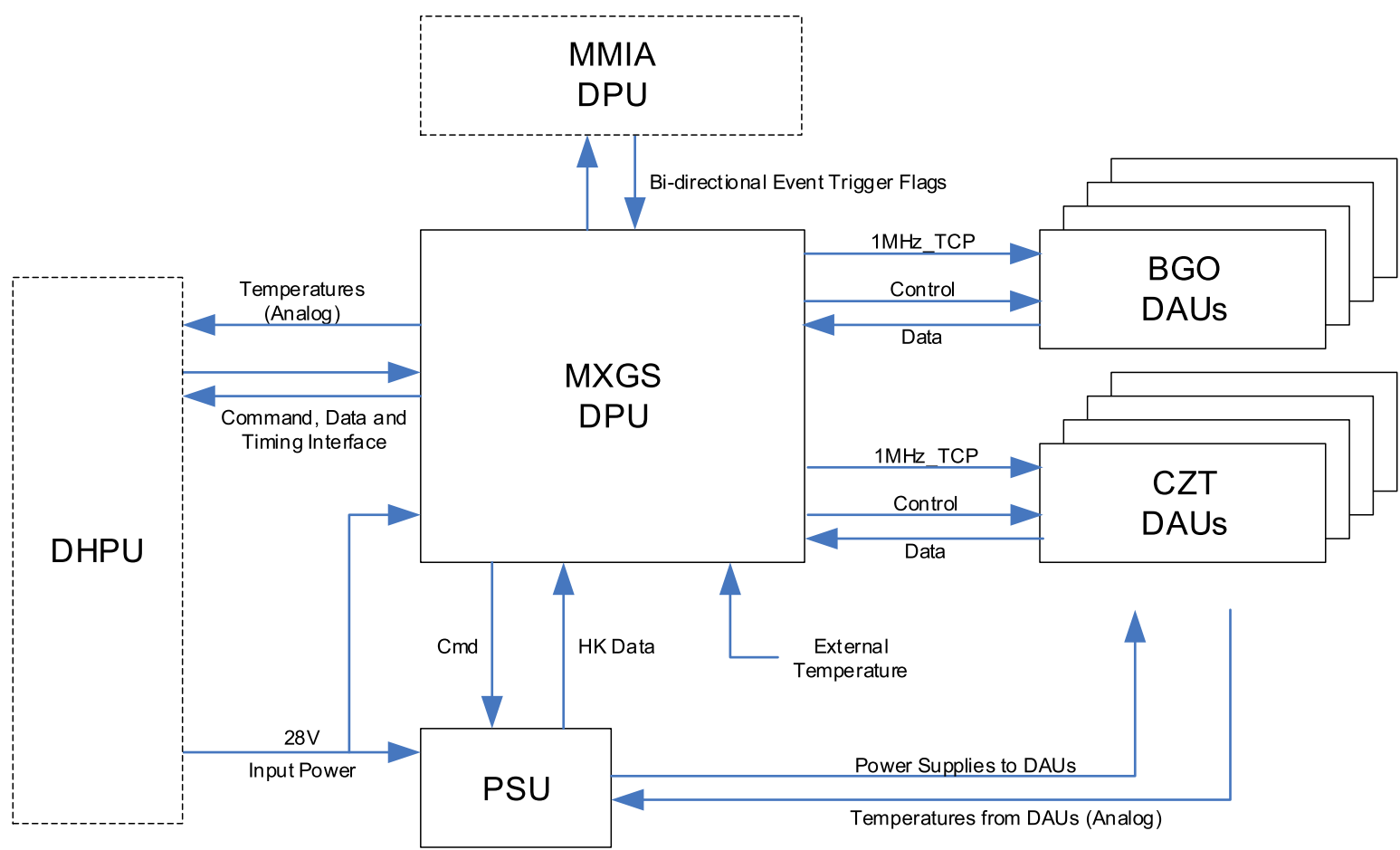

**Fig. 9** MXGS DPU top-level interface block diagram

with UTC this signal enables a relative timing accuracy of 10 µs between the MMIA and the MXGS.

The DPU also collects HK data and status data from the LED, HED and PSU and measures external temperatures of the MXGS instrument. The top-level interface diagram of the instrument DPU is shown in Fig. 9.

The DPU receives commands from the ASIM DHPU. These commands will include instrument mode change and configuration commands, data request commands and memory loads. The DPU then sends command acknowledgements back to the DHPU.

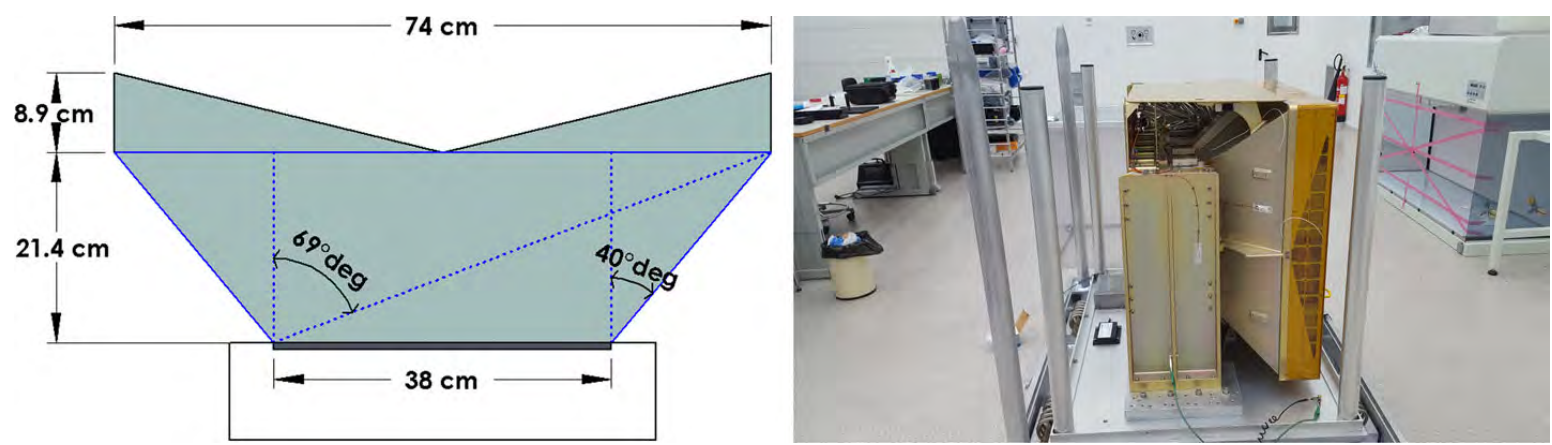

**Fig. 10** Left: The collimator mechanical envelope. Right: The collimator during integration of flight model

When requested, the DPU transmits data to the DHPU and these data will consist of both science and HK data. The data are processed and formatted by DPU for the ASIM DHPU before TM downlink.

### 3.6 The Collimator and Field of View

A hopper-shaped collimator mounted on top of the LED detector layer defines the LED Field of View (FOV) and LED Fully Coded Field of View (FCFOV). The main purpose of the collimator is to shield the LED array from X-rays coming outside the Earth horizon, called the Cosmic Diffuse X-ray Background (CDXB). The hopper walls upwards from the detector array boundary at an angle of $+-40°$ inclined from the nadir axis define the MXGS FOV of $138° \times 138°$ (Earth angular size from the ISS) and the MXGS FCFOV of $80° \times 80°$. TGFs observed within FCFOV secures an accurate imaging, while marginal TGF imaging is also possible for strong TGF at any Earth surface position. The hopper walls made in Aluminum (3 mm) and Tungsten (0.1 mm) provides a good shielding up to 60 keV (CDXB peak). Figure 10 shows the collimator geometric envelopes and the collimator flight model with the coded mask on top during MXGS integration. The collimator also has the function of shielding the detectors from Sun optical and infrared radiation. The coded mask is further described under the section "Imaging".

### 3.7 The Engineering Budget

Mass, dimension, power consumption and data rates for MXGS are as follows:

- The MXGS has a mass of nearly 147 kg and external dimensions of $(774.9 \times 746 \times 523.6)$ mm. Those dimensions include radiators, Loop Heat Pipes and Axial Groove Heat Pipes, but exclude the Multi Layer Insulation.
- The total power consumption of the MXGS, including LED, HED, PSU and DPU is $\sim 70$ W.
- The DHPU will collect up to 240 MByte of science and HK telemetry data from the MXGS per day corresponding to a mean data rate of 23 kbps.

## 4 MXGS Sensitivity and Performance

### 4.1 MXGS Performance Specifications

The MXGS performance specifications are given in Table 3.

**Table 3** MXGS specifications

| Performance parameter | Parameter value |
| --- | --- |
| Energy range | LED: 20 to 400 keV<br>HED: 200 keV to >20 MeV |
| Energy resolution | LED: <10% @ 60 keV<br>HED: <20% @ 511 keV |
| Geometrical sensitive area | LED: 1024 cm$^2$ – 16384 pixels<br>HED: 900 cm$^2$ – 12 BGO bars |
| Quant. Eff. | LED: >0.94 @ 100 keV<br>HED: >0.60 E >1 MeV |
| Relative time accuracy | <5 μs |
| Burst rate capability | LED: >350 cts/ms<br>HED: >650 cts/ms |
| Instrument background* | LED: 5–16 cnts/cm$^2$/s<br>HED: 3–12 cnts/cm$^2$/s |
| TGF sensitivity** | LED: >5–10 cnts/300 μs<br>HED: >5–10 cnts/300 μs |
| Fully Coded Field of View | LED: 80°×80°<br>HED: N/A |
| Location accuracy (STD error circle)<br>(based on RHESSI mean TGF spectrum) | Point source: <0.7°<br>3° diffuse source: <2.0° |

[a]This includes both cosmic background and instrument noise as measured during commissioning phase at all magnetic latitudes of the ISS orbits

[b]This is based on the following:
– 300 μs is the most sensitive trigger window
– the trigger threshold is based on background where instrument effects are removed by the trigger logic algorithm implemented by the onboard software
– the values depend on Trigger Logic Settings

### 4.2 Energy Response Matrix and Detection Efficiency

A full-scale detailed Geant4 model of the ASIM instrument mounted on the ISS was created to calculate energy response matrices (ERMs). The ERMs were generated by shooting mono-energetic photon beams onto the modeled detectors for all polar and azimuthal angles with a 15° resolution and then the energy deposition in LED and HED was recorded separately. Monoenergetic beams of 40 logarithmically spaced energies between 10 keV–1 MeV and 100 keV–100 MeV were used for LED and HED, respectively. By forward modeling the ERMs will be used to constrain models of the incident photon spectrum from the one measured by MXGS and thereby be used as an estimate of production altitude.

Figure 11 shows the MXGS on-axis/zenith effective area for both LED and HED, as was estimated from the corresponding ERM. The decrease of the effective area of LED above 100 keV is a consequence of the increasing transparency of the 5 mm thick CZT crystals with energy. The LED coded mask, 1 mm thick Tungsten, with support structures reduces the effective area of LED by a factor of ∼2.5. The HED effective area increases with energy as the LED and the coded mask becomes more and more transparent. The effective area of the HED becomes larger than the geometric area (900 cm$^2$) at energies greater than 20 MeV. This happens because the detector can detect secondary particles and photons due to interaction of primary photons in the surrounding materials, other payloads and the Columbus module. The ERMs are accurate to within 3% where the effective area is large, and to within 10% in the energy range with small effective area.

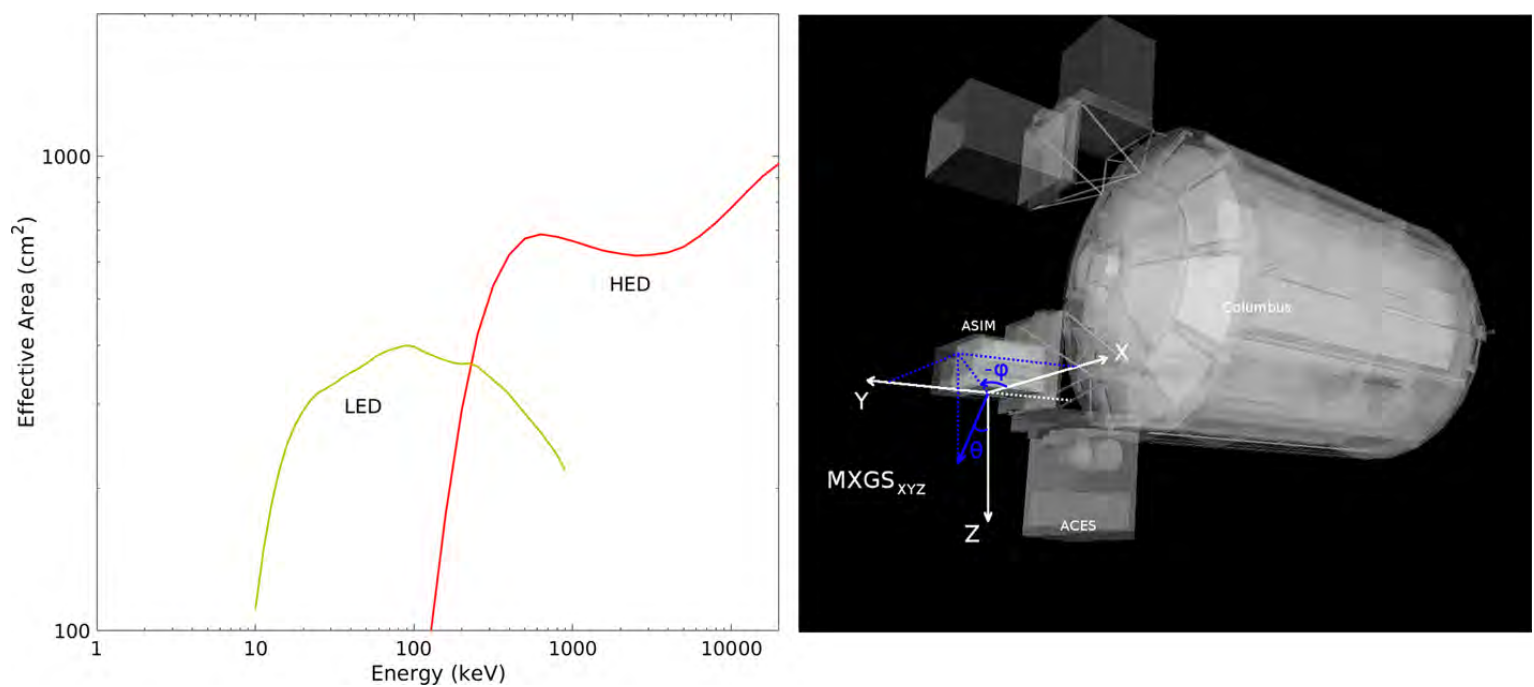

**Fig. 11** Left: The on-axis effective area of LED and HED as a function of incident photon energy. The effective area is estimated from the ERM. Right: The model used for ERM with coordinate systems for MXGS (white) and ERM (blue)

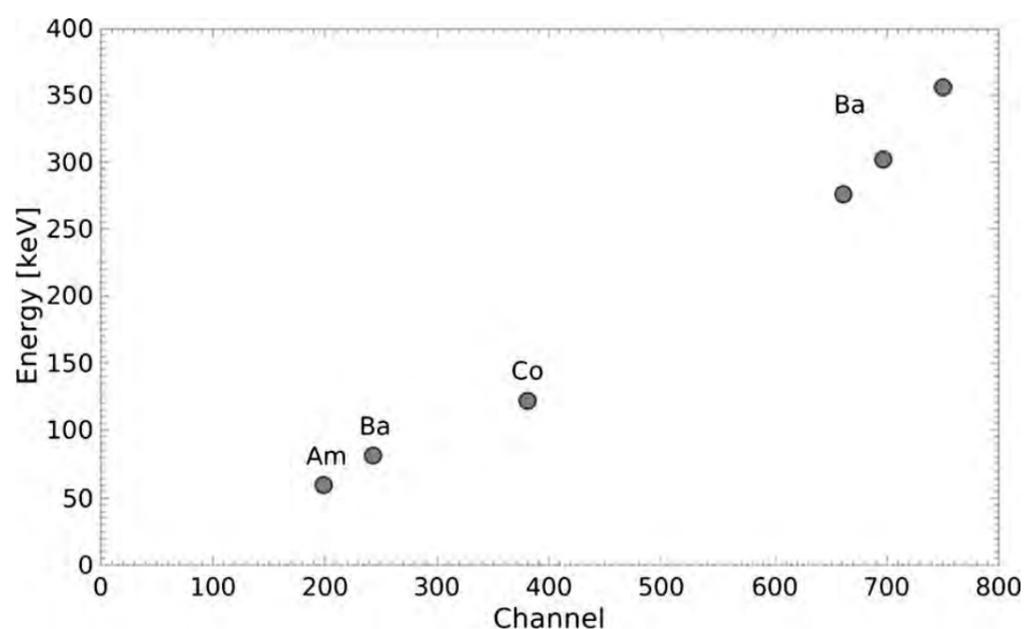

**Fig. 12** Typical $^{133}$Ba, $^{241}$Am and $^{57}$Co peak locations for a single pixel

### 4.3 Pre-Launch Energy Calibration

#### 4.3.1 LED

The pre-launch calibration of the LED has been performed using three radioactive sources: $^{133}$Ba, $^{241}$Am and $^{57}$Co.

To identify source peak locations, the data have been analyzed pixel by pixel. For 97% of the pixels four peaks of $^{133}$Ba, one of $^{241}$Am and one of $^{57}$Co sources have been identified. They are at 81 keV, 276 keV, 302 keV and 356 keV of $^{133}$Ba, 60 keV of $^{241}$Am and 122 keV of $^{57}$Co. Typical peak positions for a single pixel are shown in Fig. 12 and functional fit to these positions are used to make a channel to energy (keV) conversion for each pixel.

**Table 4** List of isotopes used for the linear calibration of the HED with their $\gamma$-lines. The 1785 keV energy is the coincidence peak between the 511 keV and the 1274 keV lines of $^{22}$Na

| Isotope | Energy, keV |
|---|---|
| $^{133}$Ba | 303 |
| $^{133}$Ba | 356 |
| $^{22}$Na | 511 |
| $^{137}$Cs | 662 |
| $^{22}$Na | 1274 |
| $^{22}$Na | 1785 |

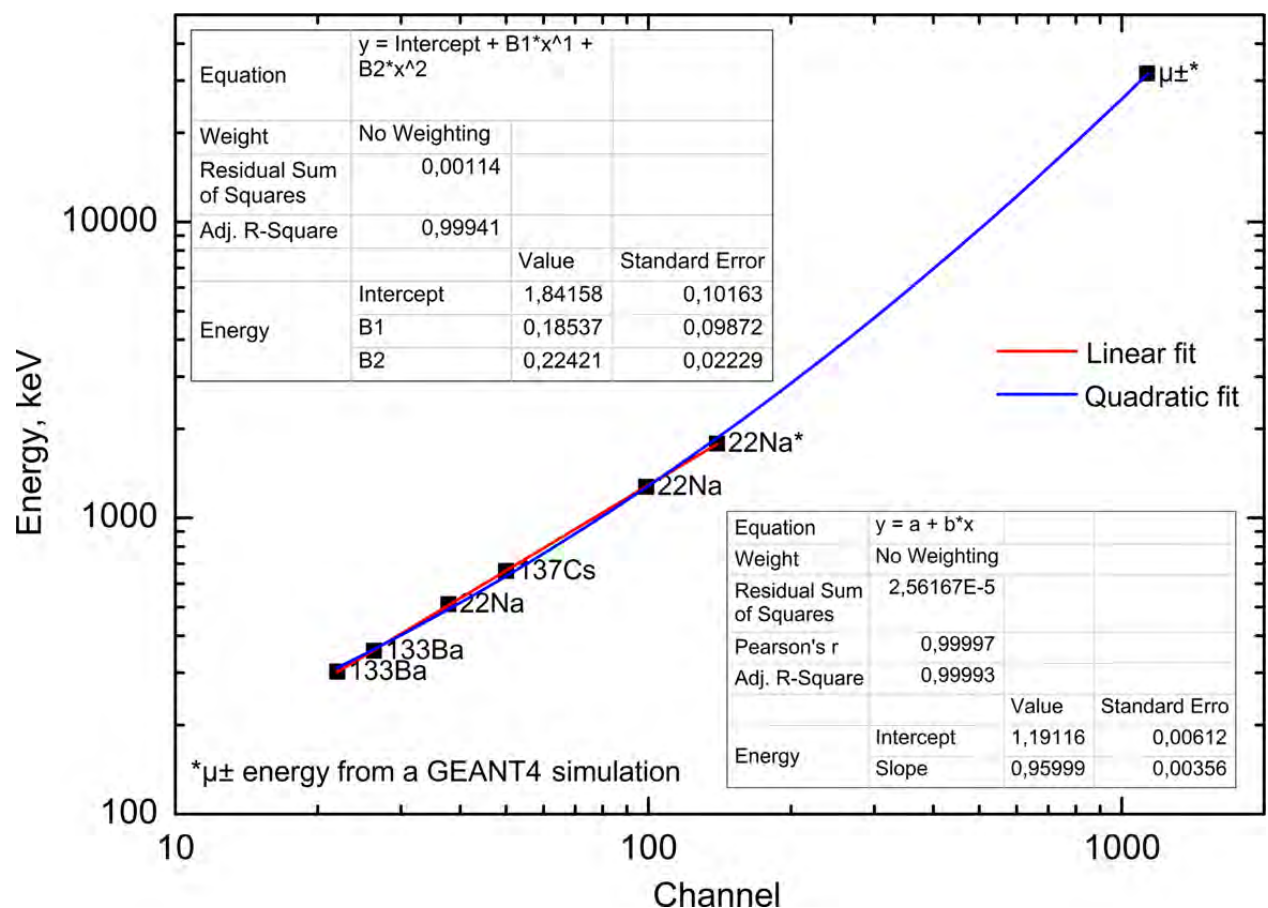

**Fig. 13** The linear calibration (red) and the quadratic calibration (blue). The muon peak seen on ground was modeled by GEANT4 to be at 31.7 MeV

#### 4.3.2 HED

As opposed to the LED, the HED performance is strongly dependent on the temperature due to the temperature dependent scintillation efficiency of the BGO. Consequently, spectra have been taken at seven different temperatures, from $-20$ °C to $+40$ °C, with 10 °C step. The high Voltage settings for the PMTs were the same as will be used in orbit. Two fits to the calibration measurements have been performed for each BGO/PMT: linear and quadratic. The former uses the sources listed in Table 4, while the latter also includes the cosmic $\mu^{\pm}$ peak from the cosmic ray background, which was clearly seen in the BGO/PMT pairs on ground. This peak has been modeled by GEANT4 to appear at 31.7 MeV energy, for the present test geometry. Figure 13 shows example of both the linear and quadratic calibration. The performance of the PMTs is sensitive to vibrations and shock and their gains can change during launch. However, once in space, the calibration will be updated based on the onboard calibration system (see next section). In space we will also use the peak produced by high-energy protons ($>1$ GeV). This peak has been modeled by GEANT4 to appear at 31 MeV for the HED geometry in space. First data from in-orbit indicate that this peak is seen in all

**Table 5** Radioactive source capsules for the on-board calibration system on MXGS

| Source ID | $^{109}$Cd (LED) | $^{22}$Na (HED) |
|---|---|---|
| Nominal activity[a] | 2 uCi (74 kBq) | 5 nCi (185 Bq) |
| Photon energy | 22 keV and 88 keV | 511 keV and 1275 keV |
| Half-life | 453 days | 950.8 days |
| Capsule type | A3236-1 | A3236-1 |
| Dimensions | ⌀ 5 mm × 5 mm | ⌀ 5 mm × 5 mm |
| Active diameter | 3 mm | 3 mm |
| ANSI/ISO rating | C.54243 | C.54243 |
| Source provider | Eckert & Ziegler Nuclitec GmbH  Gieselweg 1  38110 Braunschweig  Germany | Eckert & Ziegler Nuclitec GmbH  Gieselweg 1  38110 Braunschweig  Germany |

[a]These activity levels are specified such that sources will decay to their nominal activity at ASIM launch date

The $^{109}$Cd sources are located under the MXGS coded mask and collimated towards the LED plane (Fig. 14, left) while the $^{22}$Na sources are mounted underneath the LED towards the HED (Fig. 14, right)

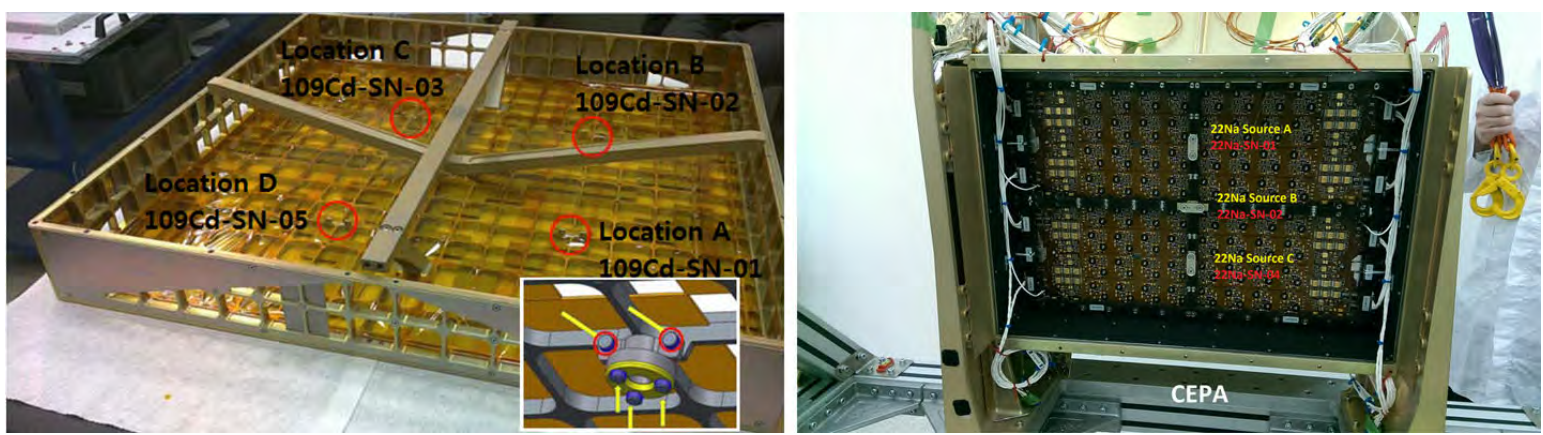

**Fig. 14** Location of sources. Left panel: Four $^{109}$Cd sources mounted underneath the coded mask radiating towards the LED. Right panel: Three $^{22}$Na sources mounted underneath LED radiating towards the HED

PMT/BGO pairs. Each PMT is protected from magnetic field effects by a Co-NETIC shield (0.3 mm thick).

### 4.4 Onboard Energy Calibration System

The MXGS has an on-board calibration system composed of four (4) collimated radioactive $^{109}$Cd sources for LED and (3) three $^{22}$Na sources for HED. The purpose of the calibration sources is to monitor the most important performance and science related parameters for the MXGS under space conditions and to update the pre-launch calibration tables. Specifications for the source capsules for the on-board calibration system are listed in Table 5.

### 4.5 Energy Resolution

#### 4.5.1 LED

The LED energy resolution estimation was performed independent from the LED energy calibration described above. The sources used here are $^{241}$Am, with a single $\gamma$-emission line

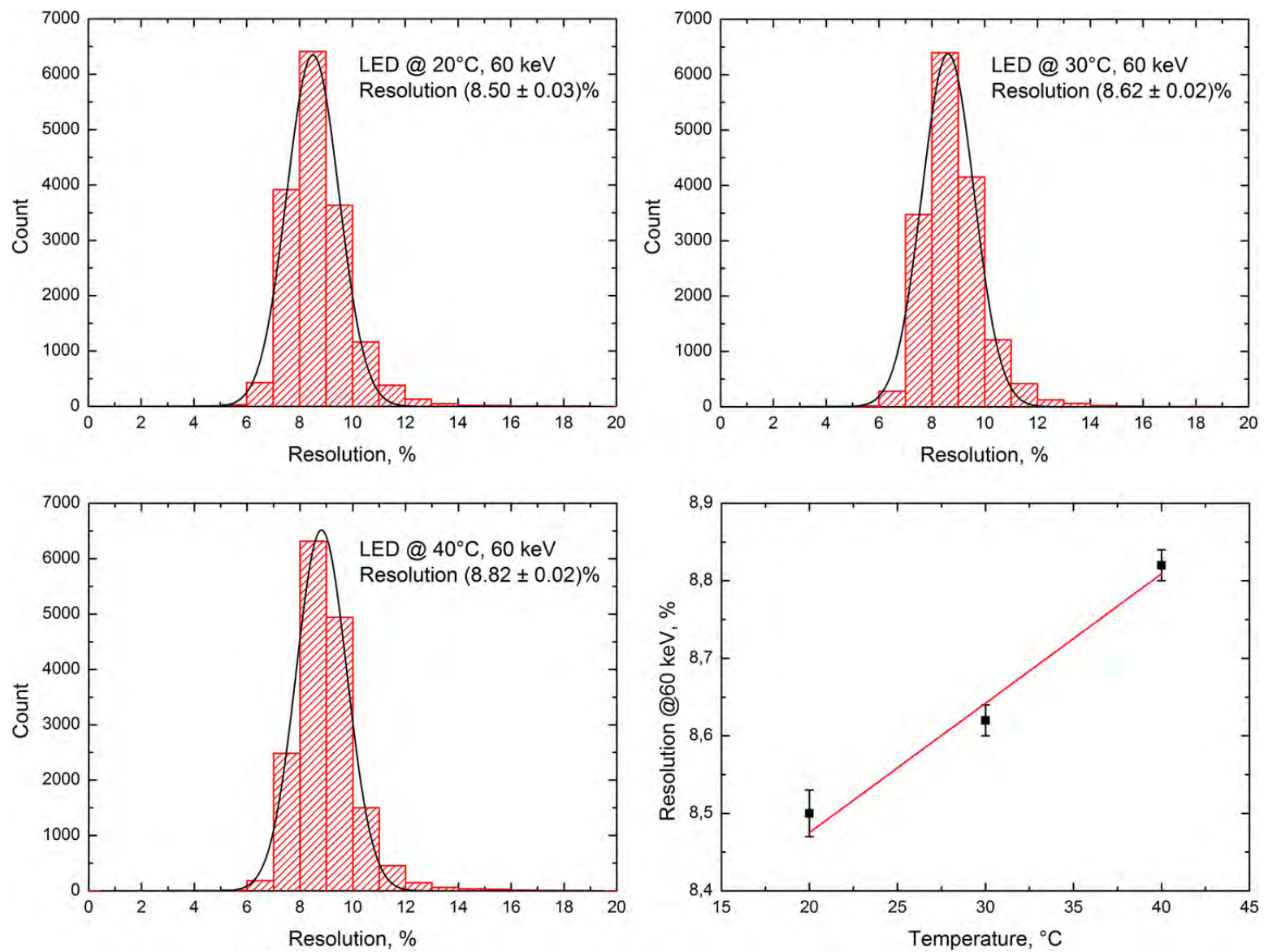

**Fig. 15** Histograms of the resolution from all healthy pixels of the LED at 20 °C (top left), 30 °C (top right), 40 °C (bottom left) and summary (bottom right)

at 60 keV, and $^{57}$Co, with a single $\gamma$-emission line at 122 keV. A linear calibration curve was assumed. The tests was performed at three different temperatures: 20 °C, 30 °C and 40 °C. In Fig. 15 we show the results for the resolution given as histograms of counts in all pixels in LED at the three different temperatures, as well as a plot showing the resolution trend versus the ambient temperature. The fractions of pixels with resolution worse than 10% are approximately 11% at 20 °C, 12% at 30 °C and 14% at 40 °C. The black curves show a Gaussian fit through the data. In the bottom-right panel is shown the dependency of the resolution on the ambient temperature. A linear fit through the points is given as a guide to the eye.

The energy range of the LED varies from pixel to pixel. Practically all pixels detect the 356 keV peak of $^{133}$Ba, meaning that an upper limit of 400 keV can be assumed for the entire LED. As far as the lower limit is concerned, relatively few pixels detect energies down to 10 keV, with the majority starting at around 30 keV.

#### 4.5.2 HED

The HED energy resolution estimation was performed independent from the HED energy calibration described above, but the high voltage settings for the PMTs were the same as will be used in orbit. Only a 1 μCi $^{22}$Na source with a single $\gamma$-emission line at 1274 keV and an annihilation peak at 511 keV was used in these tests. The resolution was defined from the 511 keV peak with an uncertainty of $\sim \pm 1\%$.

For this calibration the naturally occurring $^{40}$K (1460 keV) and $^{208}$Tl (2615 keV) peaks were used in addition to the $^{22}$Na peaks. The resolution was calculated assuming linear calibration curve, obtained from the $\gamma$-sources, without taking into account possible non-linearity at high energies. The tests were performed at three different temperatures: 20 °C, 30 °C and 40 °C. The energy resolution at 511 keV is between 15% and 20% when all pairs of PMT/BGO and all temperatures are included. The pairs of PMT/BGO were grouped such that two DAUs have approximately 200 keV to >20 MeV energy range, while the two other DAUs have approximately 300 keV to >30 MeV energy range. However, first data from in-orbit indicate that the cosmic protons, which according to GEANT4 simulation deposit a peak energy at 31 MeV, are seen in all PMT/BGOs.

### 4.6 Background Estimates

A detailed modeling of the background estimation for a Gamma Ray detector requires the knowledge of the input spectra, their fluence, type of interaction, identification of the isotope production, decay chains at different time scales (seconds to months) and accurate modeling of the mass distribution around the detector plane.

The LED background modeling was done in an early phase of MXGS development using GEANT 4.8.3 code and SPENVIS. A detailed MXGS mass model was implemented in GEANT 4, including Columbus External Payload Adapter (CEPA) and Columbus module (Alpat et al. 2010).

In a first step the prompt interactions for trapped protons and electrons (according to AP-8 MAX and AE-8 MAX models respectively); solar protons (according to JPL-91 model); cosmic protons and helium nuclei (according to CREME86, 1 conditions) and diffuse cosmic X-rays were considered. In a second step the delayed background induced by South Atlantic Anomaly (SAA) passages were included using SPENVIS for each orbit position. After exclusion of the SAA region, the prompt background rate in LED is 2.09 cnt cm$^{-2}$ s$^{-1}$, whereas the delayed background rate produces a comparable contribution for orbits having a SAA passage (4 orbits of $\sim$15 orbits daily). The cooling time defined as the time needed to reach the average background level after a SAA passage, is one orbit.

Because the MXGS mass budget slightly increased since the Phase B to the proto-flight model, our final estimation for the prompt contribution is 2.4 counts cm$^{-2}$ s$^{-1}$ in LED. For orbits after strong SAA passages it could reach 5 counts cm$^{-2}$ s$^{-1}$. Background MXGS orbital time behavior is in good agreement with the measured background in the LEGRI instrument (Sanchez et al. 1999). An estimate was also done for the HED background giving 2.0 counts cm$^{-2}$ s$^{-1}$. Comparing these values with the measured ones listed in Table 3 we will point out the following: The measured "background" contains both the cosmic background and instrumental effects. When corrected for these instrumental effects the estimated values are in good agreement at low and middle latitudes. The upper values listed in Table 3 are only measured during short intervals in orbits at very high magnetic latitudes in north Canada and south of Australia.

The background and its slow variability along the orbit is not a critical item for detection and location determination of TGFs, due to the very short duration of TGF (<1 ms). The typical background time variability is one orbit induced by the geomagnetic shielding or cooling time after a SAA passage. One-second background measurements are foreseen for each TGF trigger, so a local background estimate can be made on ground for each trigger.

### 4.7 Imaging

Imaging capability is provided by a coded mask (Fig. 16) in the front of the hopper (30.3 cm from the detector plane of LED), which leaves about half of the LED detector area illumi-nated. The geometry of the coded mask has been optimized to give the highest achievable resolution for a point source and a diffuse source of 3° Root-Mean-Square (RMSQ) beam radius. The MXGS coded mask is a 16 × 16 array of squares filled with a pattern made by the 2 × 2 cyclic repetition of an 8 × 8 Perfect Binary Array (PBA) with 46% of it closed with 46.2 × 46.2 × 1.0 mm Tungsten plates. The mask pattern is not an ideal PBA, because additional Tungsten plates have been added around the centre to perturb the cyclic nature of the pattern and to cover the mask/collimator support structure beneath. The cyclic nature of such a PBA pattern means that sources outside of a 20 × 20° central square field of view will cast an illumination pattern similar to that of a "ghost" source on the radial opposite side of the field of view. Adding more tungsten absorbing elements at the centre makes such "ghost" sources appear in an image reconstruction but with a much lower signal-to-noise ratio (SNR) so they can be distinguished from the true source.

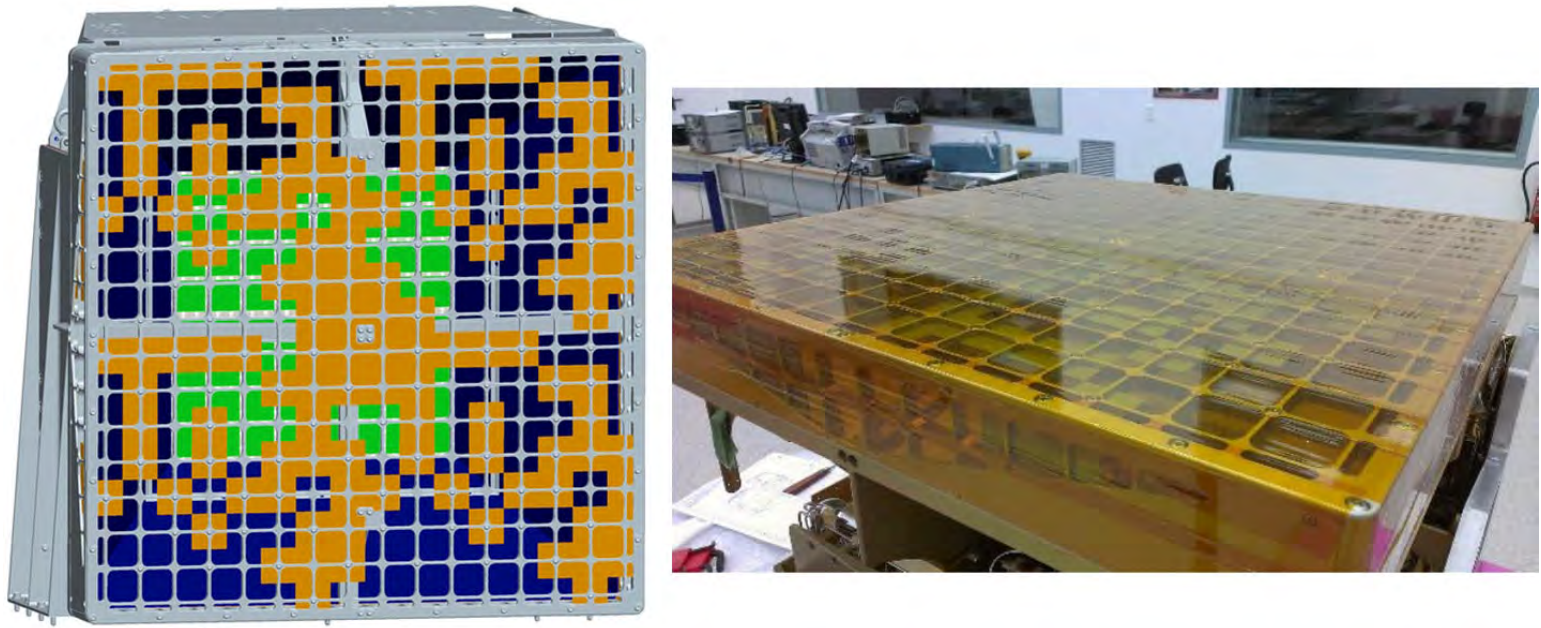

**Fig. 16** Left (a): MXGS 2 × 2 cyclic PBA coded mask pattern. Right (b): Oblique view of its aluminium support matrix

The input to TGF location imaging is a list of photon-hit locations in the 16384 pixels of the LED, each with a time stamp and energy deposited. These cover two seconds of data and contain mostly background, from which a significant millisecond TGF pulse must be extracted.

This TGF pulse will have a typically sparse LED spatial location pattern as shown in Fig. 17a. The sparse illumination pattern will correspond to a probability pattern contained in an Instrument Response Function (IRF). An example of an IRF for a TGF pulse is shown in Fig. 17b, where yellow (black) is high (low) probability for an incoming photon from a source in the FOV to be absorbed in each detector.

Thereafter a Poisson regression is employed to fit the LED count locations to those expected from the LED instrument response function. The image returned from this is a 161 × 161 array, which is a half degree resolution of the MXGS FCFOV of 80° × 80°. This image is further expanded into an 805 × 805 high resolution array, and then smoothed and processed to remove any "cyclic ghosts" and low significance artefacts. A typical result is shown in the Poisson Maximum Likelihood (ML) image of Fig. 18a and SNR image of Fig. 18b where there should remain one significant source in the ML image whose (zenith, azimuth) location can be determined.

Thereafter this location is used, with any simultaneous lightning discharge or radio event available from a lightning location network such as WWLLN, to calculate a TGF longitude/latitude/altitude (Fig. 19a). If this location data is not available a line of points (longi-tude, latitude) will be drawn on a map of the earth surface below the MXGS nadir for each possible TGF altitude in some 5–25 km range (Fig. 19b) with an uncertainty of $+/-3.7$ km.

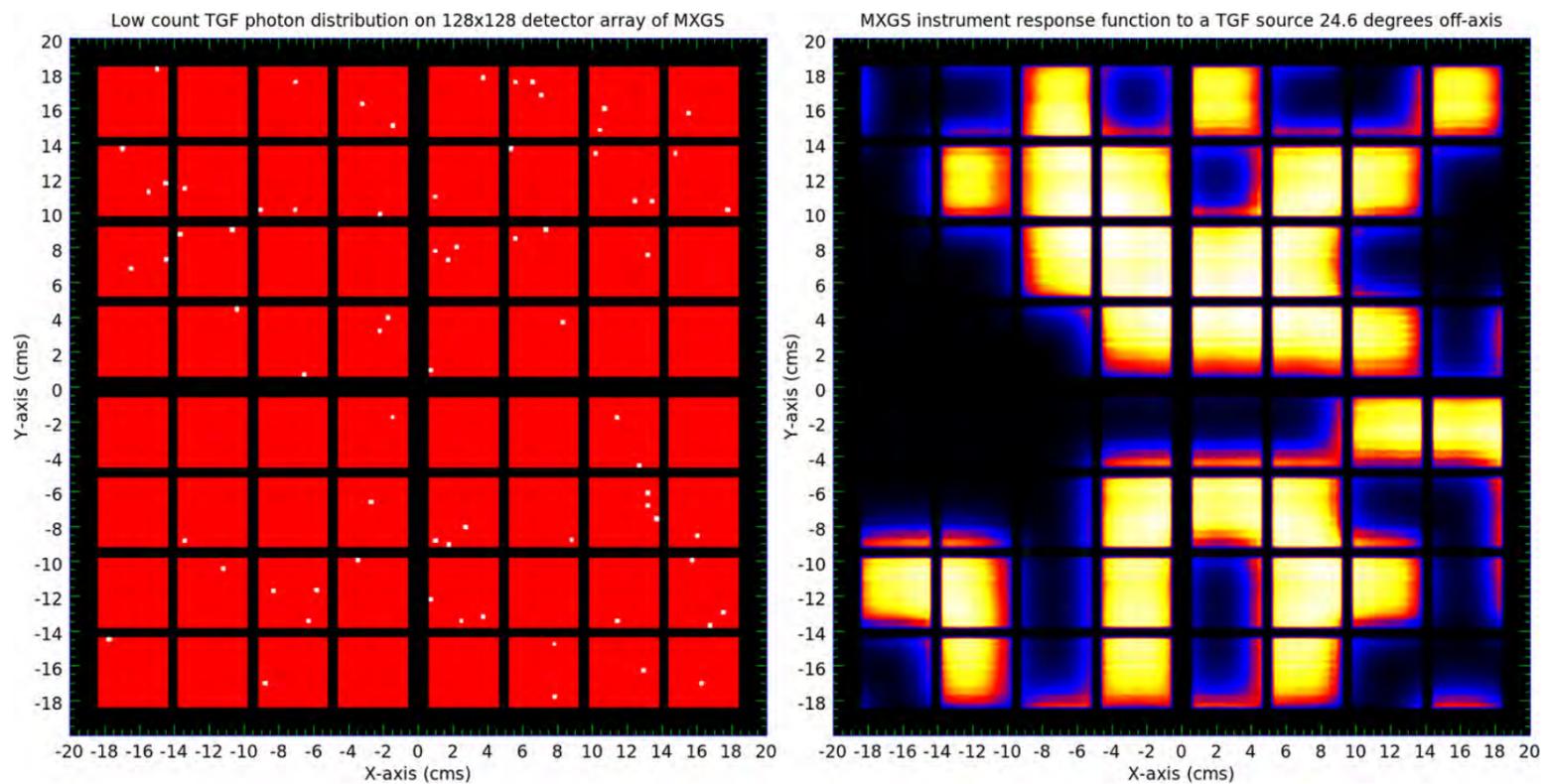

**Fig. 17** Left (a): TGF photon locations on the 16 × 16 LED module array, which is a 128 × 128 pixel array (photons are yellow). Right (b): An example of an IRF illumination pattern, where yellow (black) is high (low) probability

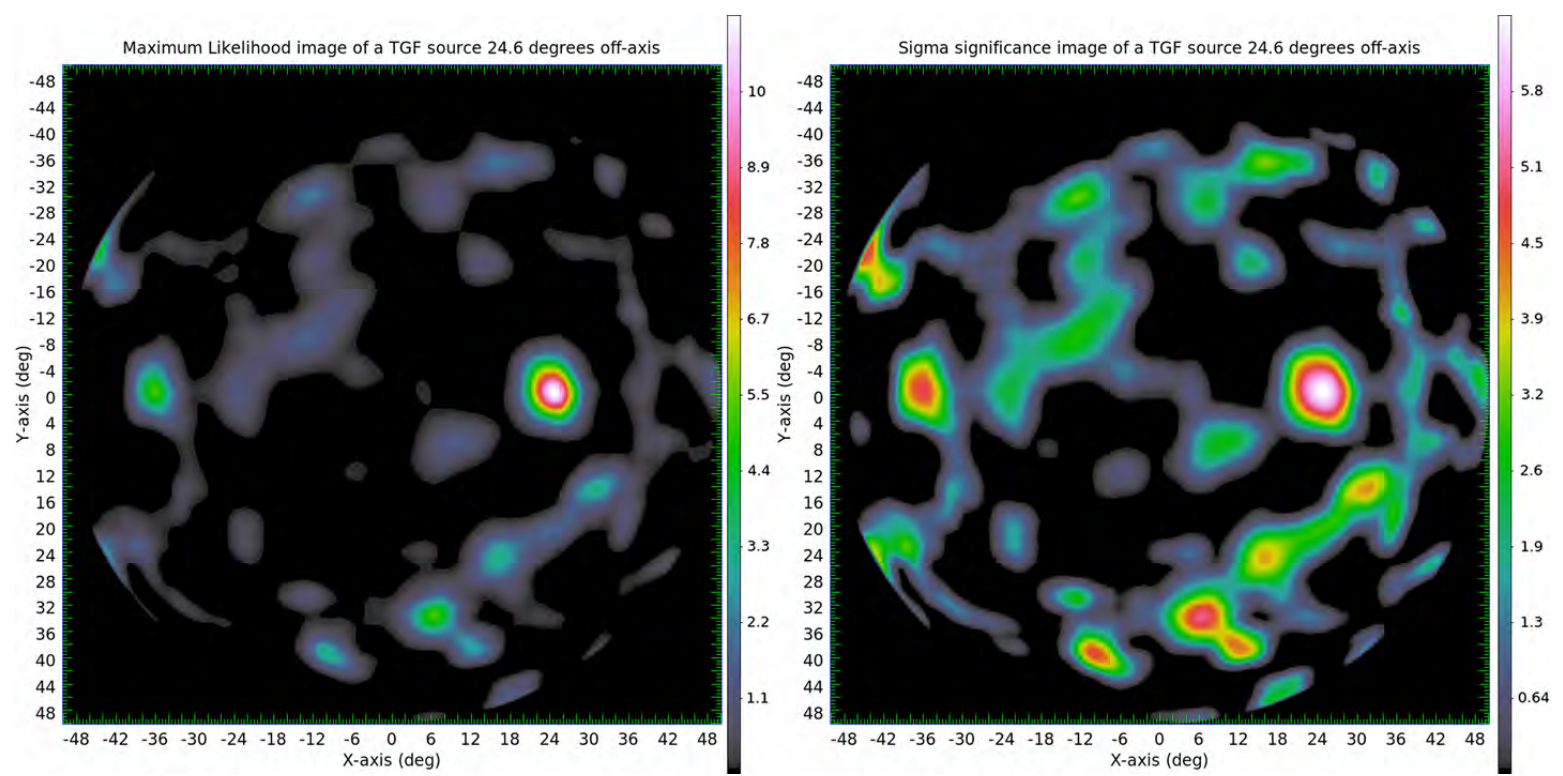

**Fig. 18** Left (a): A Poisson Maximum Likelihood (ML) image reconstruction. Right (b): The ML signal–to-noise ratio (SNR). The white spots in both panels show the most likely location of the TGF in the FOV of MXGS

With this method repeated for 500 diffuse calibration sources of Co-57, a diffuse TGF location scatter error plot has been derived, as shown below Fig. 20. The mean location error radii for point source and diffuse source are given in Table 6.

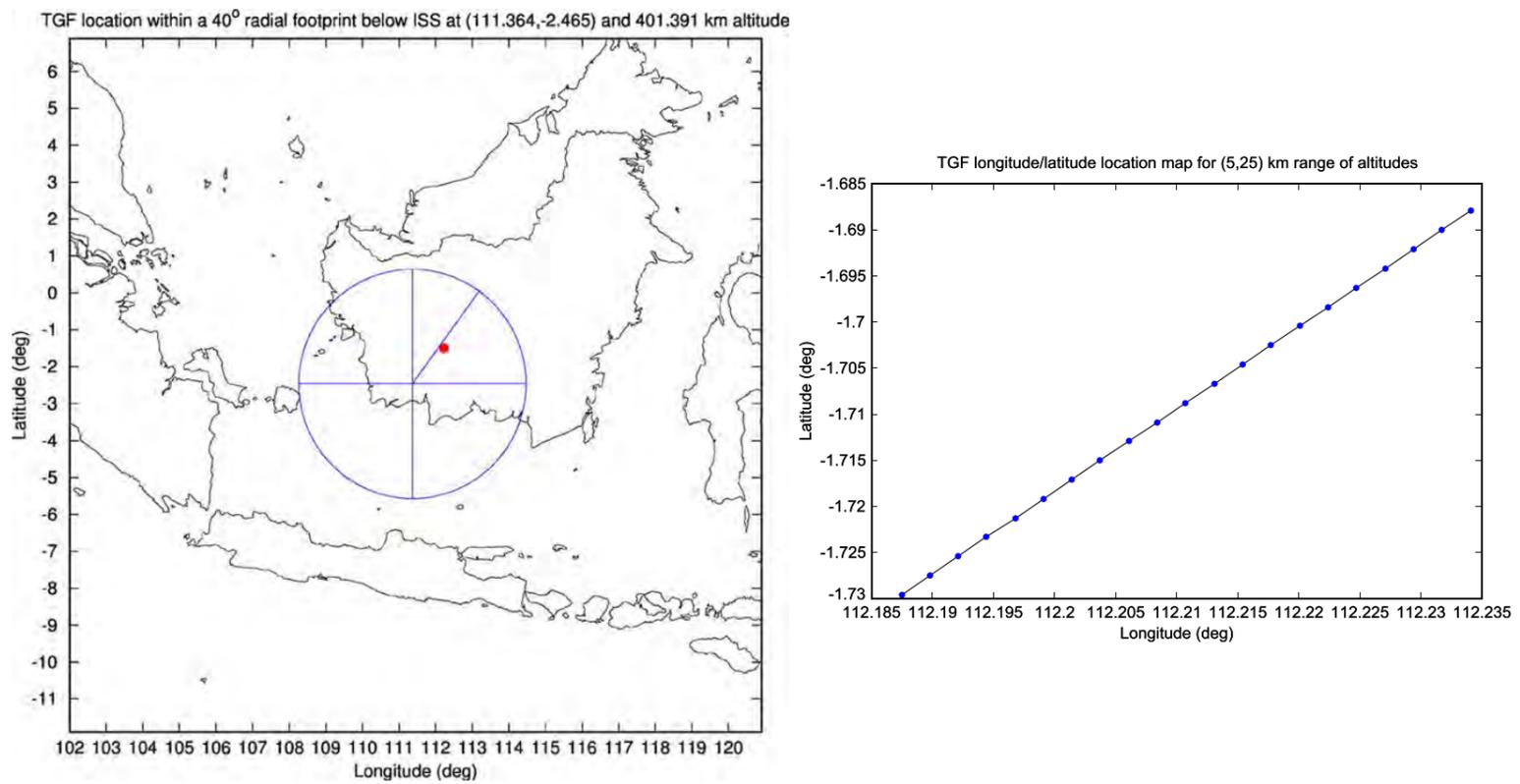

**Fig. 19** Left (a): TGF longitude-latitude map within a 45° FOV circle. Right (b): Longitude-latitude for a 5–25 km altitude range

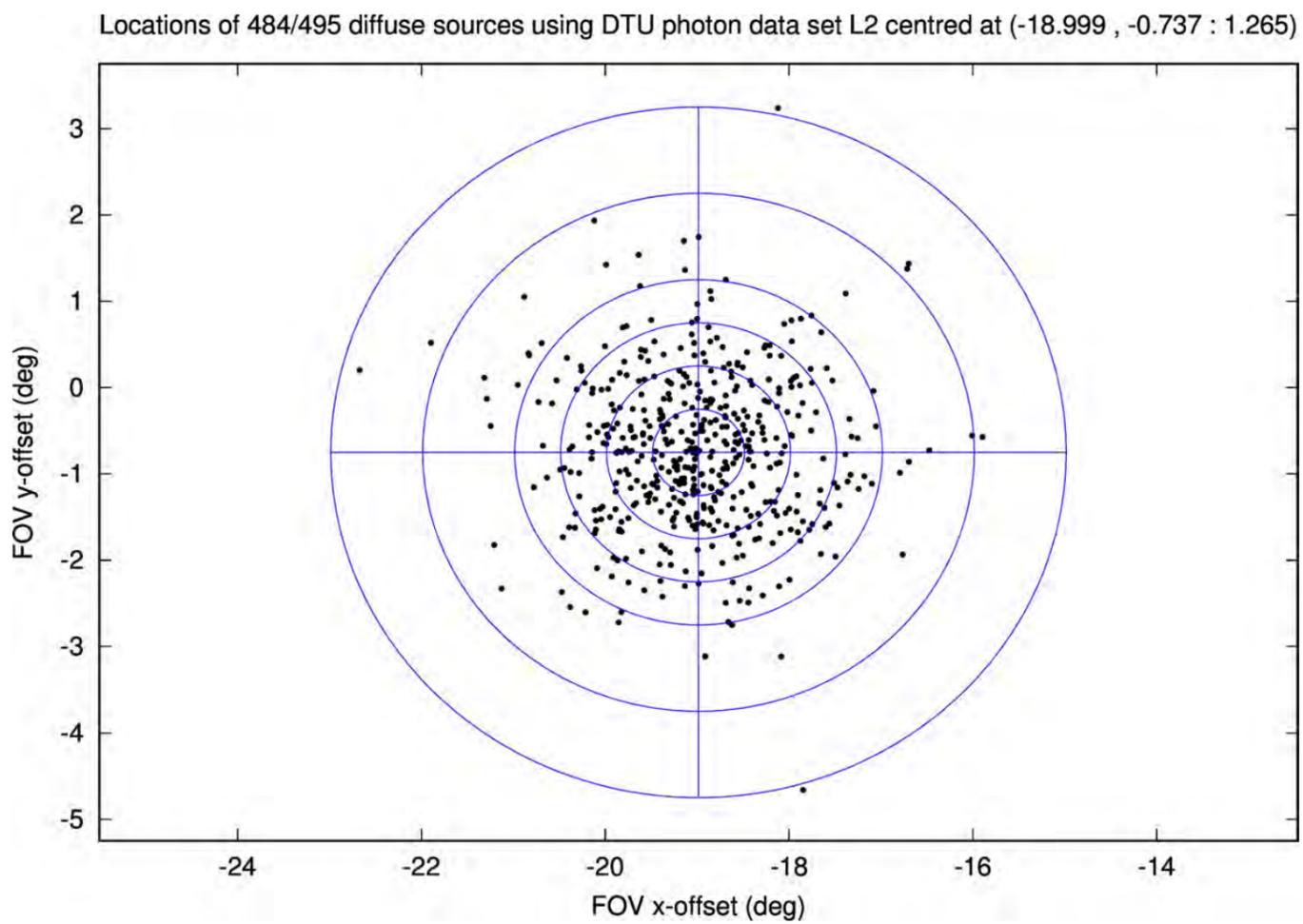

**Fig. 20** TGF location error plot for 500 C0-57 diffuse sources derive during the MXGS imaging calibration phase

**Table 6** Location error

| Pattern type PBA-8×8 | STD error circle radius (degrees) | Location confidence (%) (within 4×STD error circle radius) |
|---|---|---|
| Point source | 0.53 +/− 0.05 | 98.6 |
| Diff. source | 1.26 +/− 0.11 | 95.0 |

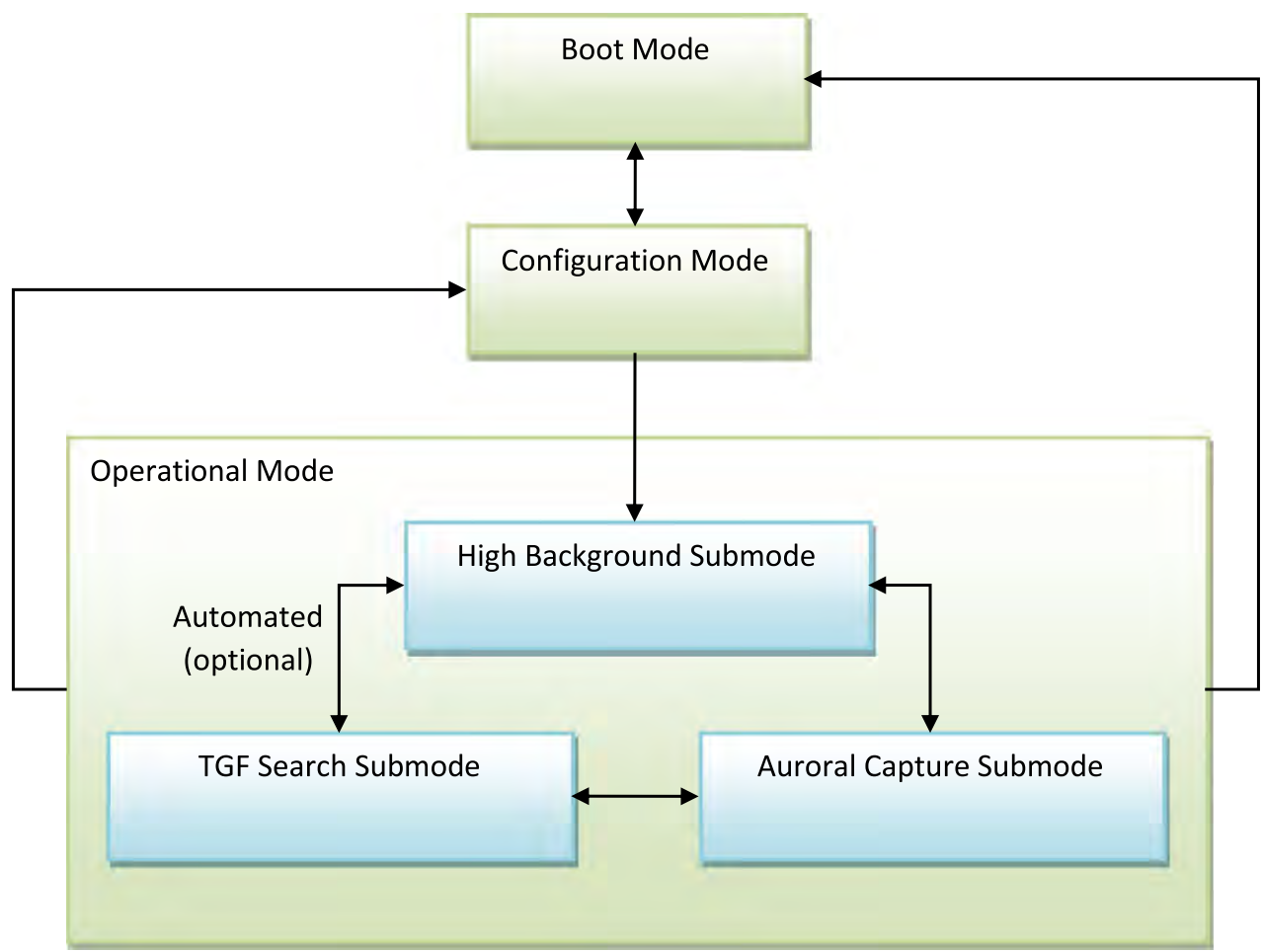

**Fig. 21** MXGS software modes, sub-modes and transitions

## 5 MXGS On-Board Software and Data Processing

### 5.1 MXGS Operating Modes and Sub-Modes

The MXGS on-board software (SW) has 3 operating modes as shown in Fig. 21.

- Boot Mode: When the MXGS is powered on, the DPU FPGA configuration is downloaded from Read-only Memory and the Boot SW is down-loaded from the electrically erasable programmable read-only memory (EEPROM). This mode supports only limited commanding and monitoring of the instrument, but includes the ability to patch the Application SW (ASW). All detectors are powered off when entering Boot Mode.
- Configuration Mode: In this mode the full capabilities of monitoring (including enabling or disabling of different types of Housekeeping (HK) data) and of control (e.g. switching on or off the different DAUs) of the instrument are available.
- Operational Mode: Once the required configuration has been established in Configuration Mode, an 'Enter Operational Mode' Tele Command (TC) is sent to MXGS to commence science observations. It is not possible to make changes to the instrument configuration in Operational Mode, although it is possible to change the monitoring, e.g. enable/disable different types of HK data, request memory dumps, etc.

Transitions back to Boot Mode may occur as a result of error conditions or may be commanded. Operational Mode has 3 science sub-modes:

- High Background Sub-mode: This sub-mode is always entered initially on transition to Operational Mode, to allow for the possibility that the transition occurs in a region of high background. The TGF Search algorithms are disabled in this sub-mode and if the DAU background count rates are high the transfer of detector count data from the DAUs to the DPU is commanded into a "grey mode" where only a fraction of detector counts are transferred, in order to avoid overloading the DPU ASW. In general, secondary science data types are generated in this sub-mode.
- TGF Search Sub-Mode: Provided the background count rate is below a specified threshold, the ASW transitions to this sub-mode and the TFG Search algorithms are enabled (autonomous transition is normally enabled by default). Subsequently, if the background count rate increases above the threshold (e.g. due to approaching the SAA), the TGF Search algorithms are disabled and the ASW returns to High Background Sub-mode.
- Auroral Capture Sub-mode: Independently of the transitions between the two above mentioned sub-modes, the ASW may be commanded into Auroral Capture Sub-mode. In this sub-mode, the ASW collects high-time resolution, pulse-height-binned background spectra directly from the DAUs, which are downlinked as Auroral Capture science observations with essentially no on-board processing. The TGF Search algorithms and collection of secondary science data are all inhibited in this sub-mode.

### 5.2 Configuration Parameter Tables

A relatively small number of Tele Commands (TCs) is defined to control MXGS, the key to flexible control of the instrument being the configuration parameter tables, which allow for configuration of all available hardware settings, together with the various parameters for the on-board data processing algorithms. Changes may be made to configuration tables in SRAM, for temporary changes, or in EEPROM, for more permanent changes.

### 5.3 TGF Trigger Algorithms and Threshold Adjustments

In TGF Search Sub-mode, the DPU ASW continually monitors the $x$- and $\gamma$-ray background, searching for TGF events. The TGF trigger algorithm searches for occurrences where the number of detector events in a defined time window $T_W$ exceed a trigger threshold of $n$ events. In practice, this is achieved by requiring that the separation between $n$ events shall be $<T_W$ to give a trigger condition.

For each detector plane, up to three short trigger windows (nominally 300 µs, 1 ms and 3 ms) and one long trigger window (nominally 25 ms) may be configured. Certain pulse-height and other criteria are applied to the "Accepted Counts" which are passed to the trigger algorithms, in order to reduce the effects of charged-particle background.

Because of the in-orbit variability of the detector backgrounds it is necessary to continuously adjust the trigger thresholds in order to maintain approximately constant false trigger rates. This is done by performing a linear fit to the previous 8 seconds of background count rates which is then extrapolated forward to be used in the next 1-second interval. Thresholds to be used for the various trigger windows are then found from look-up tables. These tables can be updated during mission operations, thereby allowing complete flexibility of threshold settings.

### 5.4 Cross Triggering Between MXGS and MMIA

When a trigger occurs within MXGS, all raw detector count data for a period of 2 seconds, approximately centred around the time of the trigger, are captured for Telemetry (TM) downlink. A hardware trigger flag is also raised to MMIA which causes MMIA to capture photometer and camera data at the time of the MXGS trigger. Similarly, when MMIA detects a trigger, e.g. due to lightning or a TLE, it raises a trigger flag to MXGS so that raw detector count data are captured for a period of 2 seconds, centred on the time of the MMIA trigger.

MXGS detector count data and photometer data within MMIA are time-stamped to a relative time precision of better than 10 µs, enabling accurate time correlation of the different scientific phenomena.

### 5.5 MXGS Science Data Types

MXGS generates 5 different types of science data observations:

- TGF Event Observation: When a TGF trigger event occurs in MXGS or when an external trigger is received from MMIA (or both), all detector raw count data for a period of 2 seconds, approximately centred on the time of the trigger, are downlinked via TM for detailed analysis on ground. This is the primary science data type.
- Background Observation: This contains pulse-height spectra of detector background with 1-second and 32-ms time resolution. It also contains values of on-board data processing algorithm parameters, e.g. trigger thresholds, pulse-height thresholds, etc. Background observations are generated at 1-second intervals but are normally bundled into 1-minute packets for TM downlink.
- Pulse-Height Spectra Observation: Background pulse-height spectra are collected over periods of nominally tens of minutes for each DAU of the LED and for each PMT of the HED, for engineering purposes including monitoring of detector gain changes.
- Sampled Detector Counts Observation: During nominal operations, 1 in every 100 raw detector counts is selected for downlink in this observation type. This is primarily to enable detailed monitoring of detector performance, in particular the evolution of hot and noisy pixels in the LED.
- Auroral Capture Observation: High-time resolution (5 ms nominal), pulse-height-binned background spectra are collected directly from the DAUs over a period of typically 10 minutes, in this mode.

### 5.6 Operation During South Atlantic Anomaly (SAA) Passages

In order to avoid ageing and consequent degradation of the PMTs in the HED DAUs due to high fluxes of charged background particles, the PMT high voltages are switched off during SAA passages, although the LED remains fully operational. The high voltage switch off is performed by the DHPU using predicted SAA boundaries. These boundaries will be updated in the light of in-orbit experience.

## 6 Summary

The MXGS instrument is the first X- and $\gamma$-ray detector specifically designed to detect TGFs. MXGS has imaging capabilities with $<0.7°$ resolution for a point source and $<2.0°$ for a diffuse source of 3° RMSQ beam radius. The MXGS instrument has two detector layers with a combined energy range of 20 keV to $>20$ MeV with a resolution of $<10\%@60\text{keV}$ and $15$–$20\%@511\text{keV}$. With fast electronics the dead-time is about 1.4 µs for LED (only applies to same chain) and 550 ns for HED. Data products from MXGS are 2 s data-strings for each triggered event in specified trigger windows, nominally set to 300 µs, 1 ms, 3 ms and 25 ms. Data-strings of 2 s will also be down-linked when MXGS receives triggers from MMIA of either TLE or lightning discharge. The ASIM mission has a lifetime requirement of 2 years.

**Acknowledgements** ASIM is a mission of ESA's SciSpace Programme for scientific utilization of the ISS and non-ISS space exploration platforms and space environment analogues. At the University of Bergen we thank for the Norwegian contribution through ESA programs. Funding at the University of Bergen, for the HED layer was supported by PRODEX contract 4000102100. Additional funding came from Norwegian Research Council (184790/V30, 197638/V30, 223252) and the University of Bergen. At DTU Space we thank the Danish Ministry of Higher Education and Science who supported ASIM from the Danish Globalization Fund for Climate Initiatives (2009-2012) via a special contribution to ESA. Development of the ASIM instrument computer was supported by the ESA PRODEX contracts PEA 4000105639 and 4000111397. The ASIM Science Data Center is supported by PRODEX contract PEA 4000115884. Additional Danish support was provided by DTU Space and Terma.

**Publisher's Note** Springer Nature remains neutral with regard to jurisdictional claims in published maps and institutional affiliations.

**Abbreviations**

| | |
|---|---|
| ADC | Analog-to-Digital Converter |
| AGILE | Astrorivelatore Gamma a Immagini LEggaro |
| ASIC | Application Specific Integrated Circuit |
| ASIM | Atmosphere-Space Interaction Monitor |
| ASW | Application Software |
| BATSE | Burst and Transient Source Experiment |
| BGO | Bismuth-Germanium-Oxide |
| CDXB | Cosmic Diffuse X-ray Background |
| CEPA | Columbus External Payload Adapter |
| CZT | Cadmium-Zink-Telluride |
| DAU | Detector Assembly Unit |
| DFEE | Detector Front End Electronics |
| DHPU | Data Handling and Power Unit |
| DM | Detector Module |
| DPU | Data Processing Unit |
| EEPROM | Electrically Erasable Programmable Read-Only Memory |
| ENOB | Effective Number of Bits |
| ERM | Energy Response Matrix |
| FCFOV | Fully Coded Field of Vie |
| FOV | Field of View |
| FPGA | Field-Programmable Gate Array |
| GBM | Gamma Burst Monitor |
| HED | High-Energy Detector |
| HK | Housekeeping |
| HVPS | High Voltage Power Supply |
| LED | Low-Energy Detector |
| LEP | Lightening-induced electron precipitation |
| LVPS | Low Voltage Power Supply |
| MMIA | Modular Multi-Spectral Imaging Assembly |
| MXGS | Modular X- and Gamma-ray Sensor |
| PBA | Perfect Binary Array |
| PCB | Printed Circuit Board |
| PMT | Photomultiplier Tube |
| PSU | Power Supply Unit |
| REP | Relativistic electron precipitation |
| RHESSI | Reuven Ramaty High Energy Solar Spectroscopic Imager |
| RMSQ | Root-Mean-Square |
| SAA | South Atlantic Anomaly |
| SCDP | SCience Data Packages |
| SNR | Signal-to-Noise Ratio |
| TC | Tele Command |
| TCP | Time Correlation Pulse |
| TGF | Terrestrial Gamma-ray Flashes |
| TM | Telemetry |
| TLE | Transient Luminous Events |